\documentclass[12pt,preprint]{aastex}

\begin{document}
\shorttitle{X-ray flares in ONC stars} \shortauthors{Getman et
al.}

\slugcomment{Accepted for publication in the Astrophysical Journal 07/17/08}

\title{X-ray flares in Orion young stars. II. Flares, magnetospheres,
and protoplanetary disks}

\author{Konstantin V.\ Getman\altaffilmark{1}, Eric D.\
Feigelson\altaffilmark{1}, Giusi Micela\altaffilmark{2}, Moira M.\
Jardine\altaffilmark{3}, Scott G. Gregory\altaffilmark{3},  Gordon
P.\ Garmire\altaffilmark{1}}

\altaffiltext{1}{Department of Astronomy \& Astrophysics, 525
Davey Laboratory, Pennsylvania State University, University Park
PA 16802} \altaffiltext{2}{INAF, Osservatorio Astronomico di
Palermo G. S. Vaiana, Piazza del Parlamento 1, I-90134 Palermo,
Italy} \altaffiltext{3}{SUPA, School of Physics and Astronomy,
North Haugh, St Andrews, Fife, KY16 9SS, Scotland, UK}

\email{gkosta@astro.psu.edu}

\begin{abstract}

We study the properties of powerful X-ray flares from 161 pre-main
sequence (PMS) stars observed with the Chandra X-ray Observatory
in the Orion Nebula region.  Relationships between flare
properties, protoplanetary disks and accretion are examined in
detail to test models of star-disk interactions at the inner edge
of the accretion disks.  Previous studies had found no differences
in flaring between diskfree and accreting systems other than a
small overall diminution of X-ray luminosity in accreting systems.

The most important finding is that X-ray coronal extents in
fast-rotating diskfree stars can significantly exceed the
Keplerian corotation radius, whereas X-ray loop sizes in disky and
accreting systems do not exceed the corotation radius.  This is
consistent with models of star-disk magnetic interaction where the
inner disk truncates and confines the PMS stellar magnetosphere.

We also find two differences between flares in accreting and
diskfree PMS stars.   First, a subclass of super-hot flares with
peak plasma temperatures exceeding 100~MK are preferentially
present in accreting systems.   Second, we tentatively find that
accreting stars produce flares with shorter durations.  Both
results may be consequences of the distortion and destabilization
of the stellar magnetosphere by the interacting disk.  Finally, we
find no evidence that any flare types, even slow-rise flat-top
flares are produced in star-disk magnetic loops.  All are
consistent with enhanced solar long-duration events with both
footprints anchored in the stellar surface.

\end{abstract}


\section{INTRODUCTION \label{introduction_section}}

A broad consensus has emerged in the past decade concerning the
structure and astrophysics of pre-main sequence (PMS) stars.  The
angular momentum inherited from the collapsing interstellar
material is mostly distributed into a rapidly rotating protostar,
a Keplerian protoplanetary disk, and frequently the orbital motion
of multiple stellar components. The young star interacts with its
disk in a complex fashion with accretion and ejection of
collimated outflow.  It is widely believed that strong magnetic
fields generated within the young star mediate this star-disk
interaction, truncating the disk and funneling accretion onto
limited portions of the stellar surface.  These issues are
developed in a variety of reviews \citep[e.g.,][]{Hartmann98,
Bouvier07a, Bouvier07b}.

It has been difficult, however, to elucidate in detail the nature
of the PMS stellar magnetospheres and their role in facilitating
star-disk interactions and accretion.  Well-developed analytical
theory assumes that a dipolar field extends to several stellar
radii where the Keplerian orbits and stellar fields are in
co-rotation, interacts with the disk magnetic field,  and guides
both accretion onto the surface and into a high-velocity
collimated outflow \citep[e.g.,][]{Shu94, Lovelace95}.  However,
considerable evidence has emerged that the magnetic fields on the
surfaces of PMS stars are concentrated in complex multipolar
active regions similar to those on the Sun, rather than a simple
dipolar.  This emerges from photometric, Doppler imaging, circular
polarization and spectroscopy of Zeeman-sensitive lines from PMS
photospheres \citep[e.g.,][]{Daou06,Yang07, Johns-Krull07,
Donati07}.  Recently, theoretical efforts have begun to calculate
the resulting complex PMS magnetosphere and accretion process
\citep{Jardine06, Gregory06a, Long07}.

X-ray studies are a potentially useful tool for investigating
these issues.  It has long been known that late-type stars exhibit
their highest levels of X-ray emission, arising mostly from
violent magnetic reconnection events, during their PMS phase
\citep[e.g., reviews by][]{Feigelson99, Feigelson07}.  Recent
X-ray surveys of nearby PMS stellar populations give detailed
insights into PMS magnetic flaring; these include the Chandra
Orion Ultradeep Project \citep[COUP,][]{Getman05} and the
XMM/Newton Extended Survey of Taurus \citep[XEST,][]{Gudel07}.
Both astrophysical study of the properties of individual flares,
and statistical study of many flares, from the COUP and XEST
observations reveal that most events are similar to solar magnetic
flaring but with X-ray luminosities enhanced $10^3-10^5$ fold in
intensity \citep[e.g.][]{Favata05, Wolk05, Flaccomio05, Stassun06,
Maggio07, Caramazza07, Arzner07, Stelzer07, Franciosini07}.

The decay of PMS X-ray flares on timescales of $10^3-10^5$~s have
proved particularly amenable to astrophysical modeling.   One
favored model, developed by \citet{Reale97} and extensively
applied to solar and stellar X-ray flares, considers the X-rays
produced by thermal plasma at $T \sim 10^7$~K confined in a
cylindrical loop cooling by radiation and conduction but subject
to possible reheating by later magnetic reconnection events. Using
such models, inferences can be made concerning magnetic fields
responsible for the flare, including loop length and magnetic
field strength.  For the most powerful PMS flares, inferred loop
lengths reach $4-20$~R$_\star$, comparable to the expected inner
edges of protoplanetary disks \citep{Favata05}\footnote{Recall,
however, that the vast majority of weaker PMS flares arise in
magnetic loops no larger than the star
\citep[e.g.][]{Imanishi03,Wolk05, Franciosini07}.}. Thus, X-ray
flaring potentially probes the region of star-disk interaction.
Links are also emerging between other features of PMS X-ray
emission, such as the statistical suppression of flares in
accreting systems and the rotational modulation of X-rays, and
models of multipolar PMS magnetospheres \citep{Gregory06,
Gregory07}.

The present study extends the analysis of powerful flares from the
COUP made by \citet{Favata05}.  They modeled 32 flares using a
traditional technique of time-resolved spectroscopy. In
\citet[][Paper I]{Getman08} and \citet{Getman06}, we introduce a
more sensitive data analysis method based on adaptively smoothed
median energies which permits modeling of 216 COUP flares.  This
is a sufficiently large sample that permits new investigation of
possible links between flaring, PMS magnetospheres and
protoplanetary disks. Here, we present observational evidence that
X-ray emitting coronal structures are in fact truncated by inner
disks around the Keplerian corotation radius, just as predicted by
PMS theoretical models outlined above.  We also report a variety
of other results, both positive and negative, linking magnetic
flare properties to the presence or absence of disks.

The paper is organized as follows.  Preliminary work and recovery
of some established results are given in \S \ref{disk_section}.
The main results of our study appear in the following three
sections: the absence of strong links between some flare
properties and disks (\S \ref{flare_prop_section}); clear evidence
that PMS magnetospheres do not extend beyond disk inner edges (\S
\ref{truncation_section}); and possible relations between
super-hot flares,  accretion and non-dipolar magnetic fields (\S
\ref{superhot_section} and \ref{magnetic_fields.section}).
Discussion follows in section \S \ref{discussion_section}. Readers
are encouraged to consult Paper I for details on the selection,
modeling, and properties of the 216 COUP flares discussed here.

\section{PRELIMINARY CONSIDERATIONS \label{disk_section}}

\subsection{Disk and accretion indicators \label{indicators_section}}

\begin{figure}
\centering
\includegraphics[angle=0.,width=6.5in]{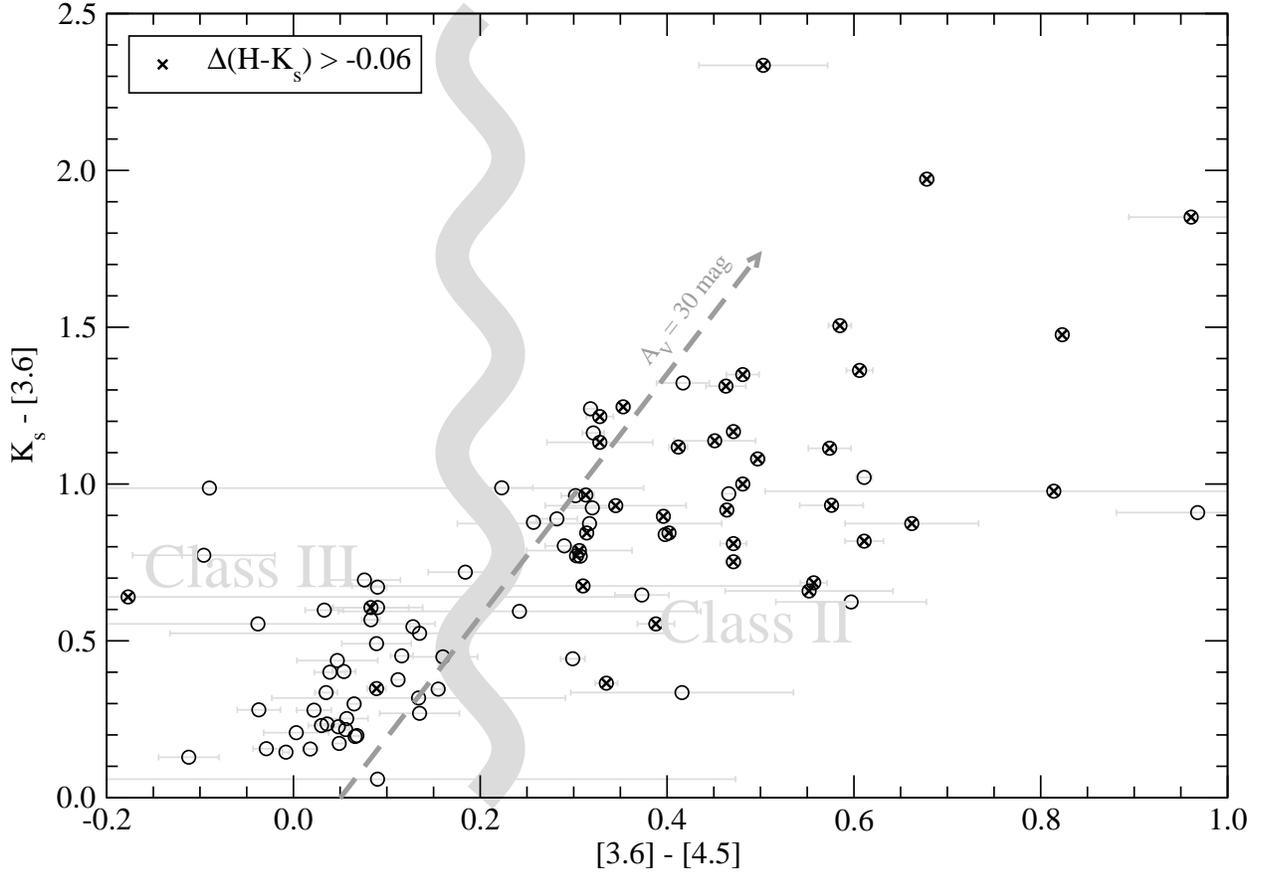}
\caption{Disk indicators $K_s - [3.6]$ $vs.$ $[3.6]-[4.5]$
color-color diagram for 98 COUP sources with available infrared
photometry. For 59 sources $\Delta(H-K_s) < -0.06$~mag indicating
no inner disk, while for 39 sources (marked by $\times$)
$\Delta(H-K_s) > -0.06$~mag indicating an inner disk. Reddening
vector of $A_V \sim 30$~mag is shown as the grey arrow. The grey
serpent line roughly discriminates between stars with and without
a MIR excess from a circumstellar disk.  Typical errors on $K_s -
[3.6]$ are 0.02~mag, while errors on $[3.6]-[4.5]$ are shown as
grey bars. \label{fig_Ksch1_vs_ch12}}
\end{figure}

We follow a well-established path in defining the presence of
disks and accretion in PMS stars. The $K_s$-[3.6]-[4.5]
color-color diagram on Figure \ref{fig_Ksch1_vs_ch12} shows 98
sources with available IR photometry (both $\Delta(H-K_s)$ and
$[3.6]-[4.5]$; see Table 4 of Paper~I). This diagram is known to
provide good discrimination between Class~I (protostellar),
Class~II (classical T Tauri) and Class~III (weak-lined T Tauri)
systems \citep[e.g.][]{Hartmann05}. The reddening vector of $A_V
\sim 30$~mag for a diskfree Vega-like spectrum assuming the
reddening law of \citet{Mathis90} demarcates the locus of
Class~III objects to the left from the Class~II and Class~I
systems to the right. All COUP 98 objects, except for three with
their color $[3.6]-[4.5]>0.5$, have their inferred X-ray column
densities $\log N_H < 22.3$~cm$^{-2}$ (Table~4 of Paper~I). For
normal interstellar gas-to-dust ratios, this corresponds to a
visual absorption of $<10-12$~mag \citep{Vuong03} and thus to the
limiting color of $[3.6]-[4.5]<0.2$ for Class~III objects. Several
stars to the left of the reddening vector and with $[3.6]-[4.5]$
color in excess of 0.2 may be systems with either higher
photometric errors and/or special reddening conditions in their
individual star-disk systems. Nevertheless, the rough mid-infrared
(MIR) color criterion of $[3.6]-[4.5]=0.2$ effectively
discriminates between Class~III and Class~II systems, and we use
this as our main MIR disk indicator. A few COUP sources with
$K_s-[3.6] \sim 2$ may be Class~I or transitional Class~I/II
systems.  One of them, COUP \#570, is classified as 0/Ib
protostellar candidate in \citet{Prisinzano07}.

$\Delta(H-K_s)$ near-infrared (NIR) excess is an indicator of a
heated inner dusty circumstellar disk.  It is measured from the
reddening vector on $J-H$ versus $H-K_s$ diagram of COUP sources
\citep[e.g. Figure 5$a$ in][]{Favata05} adopting photospheric
colors associated with PMS stars at age 1~Myr using the models of
\citet{Siess00}, and considering a reddening vector applied to
$0.1$~M$_{\odot}$ stars. Only four sources in our sample have
masses $M<0.2$~M$_{\odot}$, but 40 have $0.2<M<0.4$~M$_{\odot}$.
In order to allow better discrimination of inner disks, we can
relax the criterion $\Delta(H-K_s)=0$~mag to the value in the
range of $(-1,-0.06)$~mag corresponding to reddening vectors at
the mass range of $0.2-0.4$~M$_{\odot}$.

Using this NIR color excess measure, we find that a criterion of
$\Delta(H-K_s) = -0.06$~mag is a good discriminator between
diskfree and disky stars established with the MIR color criterion
of $[3.6]-[4.5]=0.2$; 95\% of sources classified as Class~III
using this MIR color criterion are also Class~III using our NIR
color criterion (Figure \ref{fig_Ksch1_vs_ch12}).  About 17 stars
are classified as Class~II systems using MIR colors but Class~III
using NIR colors;  these systems likely have evolved disks with
inner holes.

\begin{figure}
\centering
\includegraphics[angle=0.,width=6.5in]{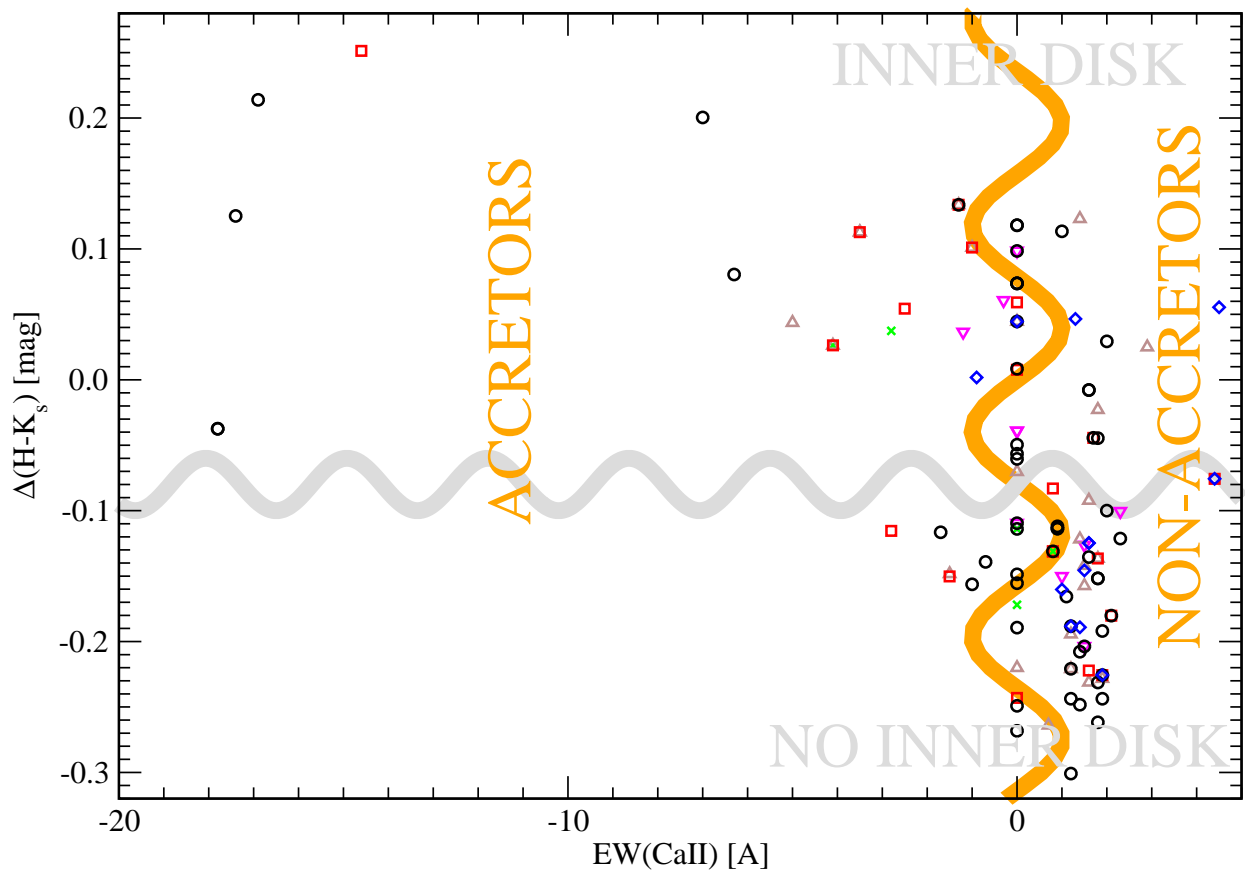}
\caption{Inner disk indicator $\Delta(H-K_s)$ $vs.$ accretion
indicator from the 8542~\AA\ Ca~II line for 128 flares on 96 COUP
stars with available data. Symbols represent flare morphologies:
57 ``typical'' flares (black circles); 21 ``step'' flares (red
boxes); 12 ``slow-rise-and/or-top-flat'' flares (blue diamonds);
23 ``incomplete'' flares (brown triangles); 5 ``double'' flares
(green $\times$); and 10 ``other'' flares (magenta triangles). The
grey serpent line roughly discriminates between stars with and
without inner disk. The orange serpent line very roughly
discriminates between stars with and without accretion.
\label{fig_dhks_vs_ewca}}
\end{figure}

Discrimination between accreting and non-accreting objects employs
the equivalent width of the IR 8542~\AA\ Ca~II line, ${\rm
EW(Ca~II)}$, measured from low-resolution spectroscopy by
\citet{Hillenbrand97}.  In Figure \ref{fig_dhks_vs_ewca}, we adopt
the classification criterion used by \citet{Flaccomio03}: stars
with the line in emission with ${\rm EW(Ca~II}) < -1$~\AA\ are
accretors while stars with the line in absorption and equivalent
width of ${\rm EW(Ca~II}) > 1$~\AA\ are considered to be
non-accretors. Stars with intermediate values $-1 < {\rm
EW(Ca~II}) < 1$~\AA\ have an indeterminate classification. Figure
\ref{fig_dhks_vs_ewca} shows that this ${\rm EW(Ca~II)}$ accretion
indicator agrees in most cases with the $\Delta(H-K_s)$ NIR inner
disk indicator: inner disk photometric excess is seen in 15 of the
19 accretors and no photometric excess is seen in 32 of the 41
non-accretors. COUP flaring sources with their conservatively
chosen ${\rm EW(Ca~II}) < -2$~\AA\ will be further classified in
the text as high-accretors\footnote{The association between
accreting stars and MIR-excess stars in our sample is not perfect.
Four high accretors located at the outskirts of the $Chandra$
field are not part of the MIR disk sample because they lie outside
the $Spitzer$ IRAC fields from which we obtained MIR photometry.
An additional high accretor lies projected against the
infrared-bright BN/KL region which is classified as a diskfree
star based on its (possibly erroneous) MIR photometry. See also \S
\ref{cloud_section} and Figure \ref{fig_ultrahot_3}.
\label{footnote_accretors}}.

Recall that photometric and inferred properties (such as age, mass
and rotation) for these COUP stars are tabulated by
\citet{Getman05}, and tables in Paper~I present various observed
and inferred stellar and flare quantities. Tables
\ref{tbl_correl_known} and \ref{tbl_correl_new} of the current
paper give general statistical properties of those quantities.
They also provide probabilities $P_{KS}$ from two-sample
univariate Kolmogorov-Smirnov tests comparing the distributions of
various quantities in diskfree and disky stars. Significant disk
effects may be present when $P_{KS} \la 0.05$. This is relevant
for such quantities as rotational period, Keplerian corotation
radius, peak flare plasma temperature, coronal loop size relative
to stellar radius, and X-ray coronal extent relative to corotation
radius.

\subsection{Disks and stellar rotation \label{rotation_section}}

Using our sample of X-ray bright COUP stars, we confirm the
well-established result that Orion Nebula Cluster PMS stars with
disks rotate slower than those without disks
\citep[e.g.,][]{Herbst02, Rebull06}. Our rotational periods are
obtained from published sources but our MIR photometry was derived
independently from {\it Spitzer Space Telescope} data as described
in Paper~I and our classification of disky versus diskfree stars
was determined as described above.

\begin{figure}
\centering
\includegraphics[angle=0.,width=6.5in]{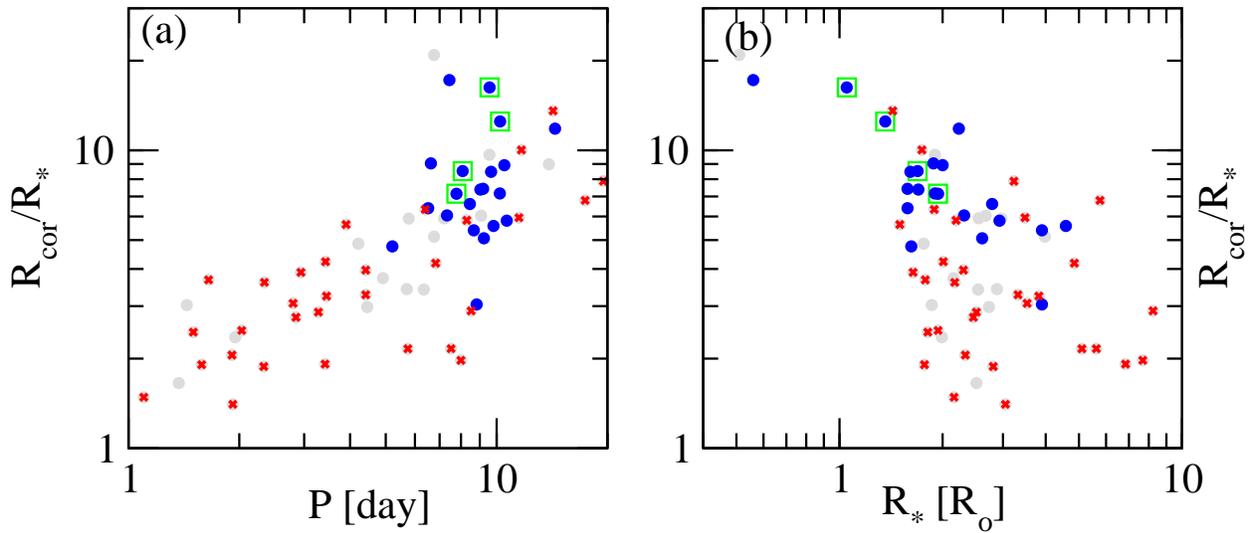}
\caption{Confirmation of PMS disk-rotation relationship: (a)
Keplerian corotation radius scaled to stellar radius $vs.$ stellar
rotational period; and (b) Keplerian corotation radius scaled to
stellar radius versus stellar radius. Panels present results for
all stars with available stellar properties (67 out of 161).
Symbols indicate subsamples: stars with MIR excess disk stars
(blue circles), without MIR disks (red crosses), uncertain disks
(grey circles), and highly accreting stars with $EW(CaII) <-2$~\AA
(green boxes). \label{fig_rotation}}
\end{figure}

This expected result is shown in Figure \ref{fig_rotation} where
stars with MIR disks (blue circles) and accretion (green boxes)
systematically have longer periods and larger corotation
radii\footnote{Recall that Keplerian corotation radii for stars
with known rotational periods $P$ and masses $M$ are calculated in
Paper~I as $R_{cor} = (G M P^2/4 \pi^2)^{1/3}$.} than diskfree
stars (red circles). This supports the explanation that slow PMS
rotation is due to the loss of a stellar angular momentum through
a magnetic star-disk interaction \citep[see review
by][]{Bouvier07a}. Table \ref{tbl_correl_known} shows that the
rotational periods of our bright X-ray flaring stars are
well-separated with a median of 9.0 days for those with MIR disks
compared to 3.5 days for those without MIR disks. Rotation of our
accreting systems are not distinguishable from other stars with
MIR disk.  It is useful to note that the range of COUP rotational
periods of $[1-10]$~days translates into the range of corotation
radii of $[2 - 10]$~R$_{\star}$ with the majority of COUP stellar
radii in $[1.4 - 3]$~R$_{\odot}$ range (Figure
\ref{fig_rotation}).

\subsection{Disks and location in the cloud \label{cloud_section}}

\begin{figure}
\centering
\includegraphics[angle=0.,width=6.5in]{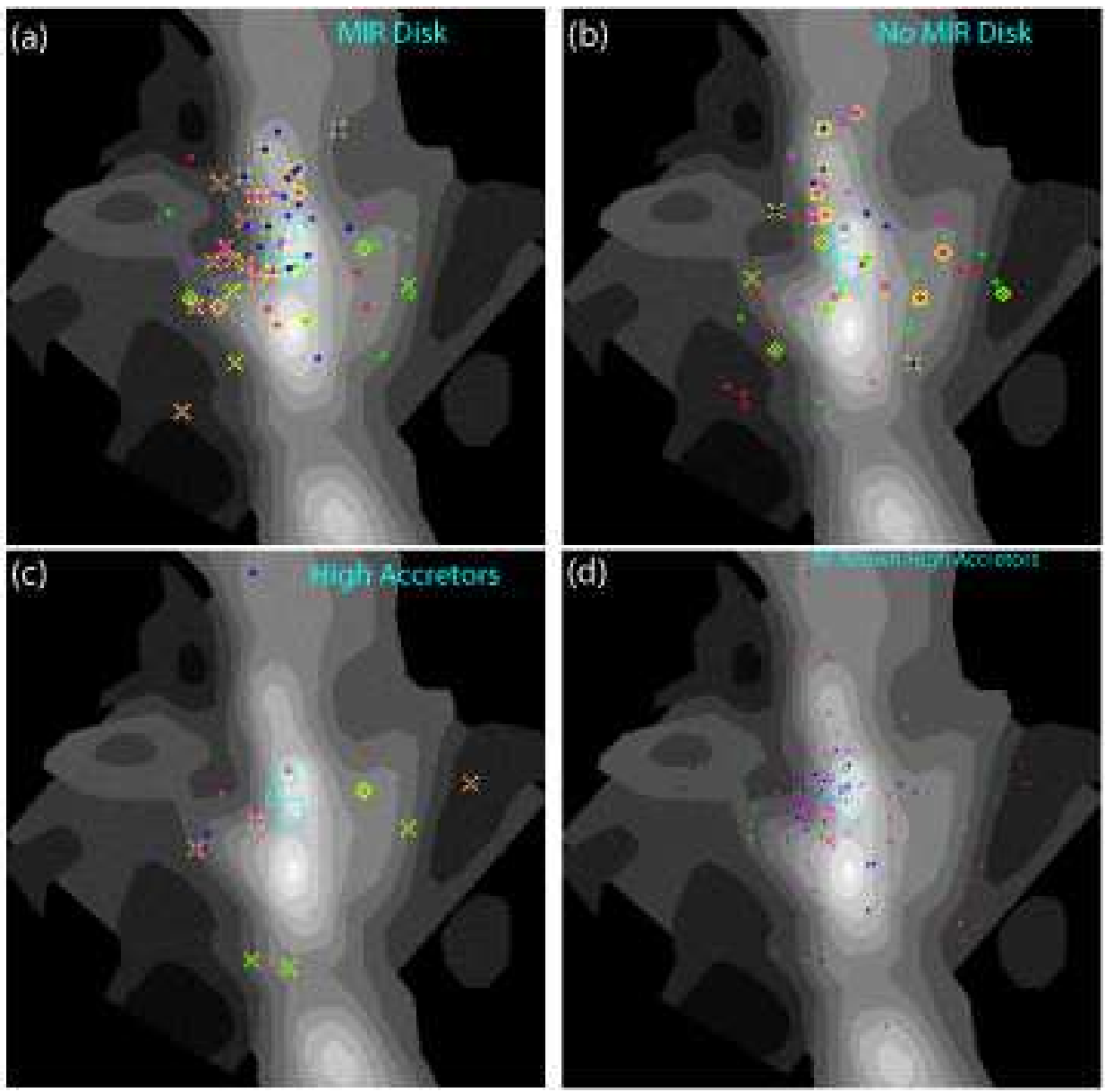}
\caption{Locations of subsamples of X-ray bright flaring COUP
stars plotted on the map of the OMC derived from the
velocity-integrated intensity of $^{13}$CO \citep{Bally87}.
Symbols are coded by their X-ray absorption:  $\log(N_H)<21$~
cm$^{-2}$ (red), $21<\log(N_H)<21.5$~cm$^{-2}$ (green),
$21.5<\log(N_H)<22$~cm$^{-2}$ (magenta), and $22<\log(N_H)<
22.8$~cm$^{-2}$ (blue). Stars with masses $M>2$~M$_{\odot}$ are
outlined by yellow diamonds; stars harboring super-hot
($T_{obs,pk} > 100$~MK) flares are labelled by yellow $\times$.
Two small cyan boxes indicate BN/KL and OMC1-S regions
\citep{Grosso05}, and the cyan circle marks $\theta^{1}$~Ori~C at
the center of the Orion Nebula Cluster. \label{fig_ultrahot_3}}
\end{figure}

Figure \ref{fig_ultrahot_3} shows the spatial distribution of our
stars with and without MIR disks in the Orion Nebula region. The
region is complicated: the rich, optically bright Orion Nebula
Cluster has low absorption and lies in front of the two OMC 1
molecular cloud cores, as well as widely distributed molecular
cloud material, where PMS stars are highly absorbed.  In the
figure, the stars are coded both by their disk properties and by
their absorption measured from the soft X-ray absorption in their
COUP spectra \citep{Getman05}. The spatial distribution of the
COUP stars is further discussed by \citet{Feigelson05} and
\citet{Prisinzano07}.

Bright flaring COUP stars with MIR disks and the highest
absorption (blue circles in panel $a$) are largely concentrated
around OMC 1 molecular filament to the north of the
Becklin-Neugebauer star forming region. Those with intermediate
absorption have a somewhat broader distribution centered
north-east of the OMC cloud cores. The diskfree stars in our
sample are generally less absorbed than those with MIR disks with
a dispersed spatial distribution similar to the disky stars with
intermediate absorption (Figure \ref{fig_ultrahot_3}$b$). For both
samples, stars harboring super-hot ($T_{obs,pk} > 100$~MK, see
\S~\ref{superhot_section}) flares are localized within the
north-eastern part of the cloud. The high accretors appear widely
dispersed (Figure \ref{fig_ultrahot_3}$c$), but this is a
selection effect - only 12 high accretors have their observed
X-ray net counts above the count threshold of our flare analysis
($NC = 4000$~counts, see Paper~I and footnote
\ref{footnote_accretors}). The total known population of COUP high
accretors have spatial distribution similar to that of the bright
flaring COUP MIR disk stars with intermediate absorption (Figure
\ref{fig_ultrahot_3}$d$).

\section{FLARE PROPERTIES AND PROTOPLANETARY DISKS \label{flare_prop_section}}

Paper I presents in detail our analysis of the selected COUP flares and the
derivation of flare properties used in the analysis here: rise and decay
timescales, ``characteristic'' pre-flare and peak flare X-ray luminosities, peak
flare plasma temperature and emission measure, flare morphology, and
magnetic loop length responsible for the flare.  These
quantities are derived using the astrophysical model of flare decays developed
by \citet{Reale97}. This is a single loop model and is simplistic in a number of
ways (see its limitations in sections 2.2 and 2.5 of Paper~I). However, even in
the case of complex flares, it is appropriate to apply the model to a lightcurve
segment if there is an indication for the presence of a single ``dominant''
flaring structure (see sections 2.5 and 2.6 of Paper~I).

Adopting the model of Reale et al., \citet{Favata05} have analysed the strongest 32 COUP
flares using the long-standing flare analysis method of time-resolved
spectroscopy (TRS). To extend the flare sample of \citet{Favata05}, Paper~I
utilizes a more sensitive technique of flare analysis, the ``method of
adaptively smoothed median energy'' (MASME), introduced by \citet{Getman06}.
Instead of performing classical TRS with XSPEC over only a few characteristic
flare intervals, we employ an adaptively smoothed estimator of the median
energy of flare counts and count rate to infer the evolution of plasma
temperature and emission measure at dozens of time points along the decay phase
of a flare. This is achieved by calibrating median energies and count rate to
temperatures and emission measures through simulations of high
signal-to-noise spectra at fixed source's column density. The method permits
modeling of flare spectra on more rapid timescales  and with weaker signals
than was possible using traditional spectral fitting. Readers are encouraged to
consult section 2.2 and Appendices A and B of Paper~I on details of the method.
The result is that Paper~I  characterizes 216 COUP flares for analysis here,
in contrast to 32 flares analyzed by \citet{Favata05}.

\subsection{Disks have no effect on flare morphology \label{morphology_section}}

Paper I describes our qualitative classification of X-ray flares
by their lightcurves during the flare.  The four classes are:
`typical' flares with fast rises and slower decays characteristic
of most solar flares; `step' flares with extra emission during the
decay attributable to a reheating or a secondary reconnection
event;  `double' flares with two peaks suggestive of two nearly
simultaneous reconnection events; and `slow rise, top flat' (SRTF)
flares.  The first three classes are commonly seen in solar
flares.  For example, Figure~11 of Paper I shows solar flares with
secondary events during the decay of powerful long-duration
events.  Such events, scaled up several orders of magnitude, could
be classified as step or double flares in COUP lightcurves.

The SRTF flares are more unusual, and one may speculate that they
may be selectively formed in certain PMS stars.  For example, they
might be reconnection events associated with sheared star-disk
magnetic fields \citep{Montmerle00, Isobe03} rather than events
associated with field lines attached to the star. Flare morphology
is coded by different symbols in Figures \ref{fig_dhks_vs_ewca},
\ref{fig_L_vs_dHKs}, \ref{fig_L_vs_ch12} and \ref{fig_L_vs_ewca}.
We examine here the distribution of flare types along the abscissa
measuring the strength of accretion, NIR inner disk, and MIR disk
indicators. No pattern between flare morphology and disk
indicators is seen. In paricular, SRTF flares are seen in systems
with and without MIR and NIR disks.  However, SRTF flares are not
seen in high-accretion systems; this can be an indication of a
real physical effect (\S \ref{energetics_section}) or simply
attributed to the very limited statistics of both accreting and
SRTF flare samples.

\subsection{Disks are unrelated to flare energetics \label{energetics_section}}

We examine here whether any relationships are present between
disks and quantities associated with the strength of the X-ray
flares:  rise and decay timescales, peak luminosities, and total
energies in the X-ray bands.

\begin{figure}
\centering
\includegraphics[angle=0.,width=6.0in]{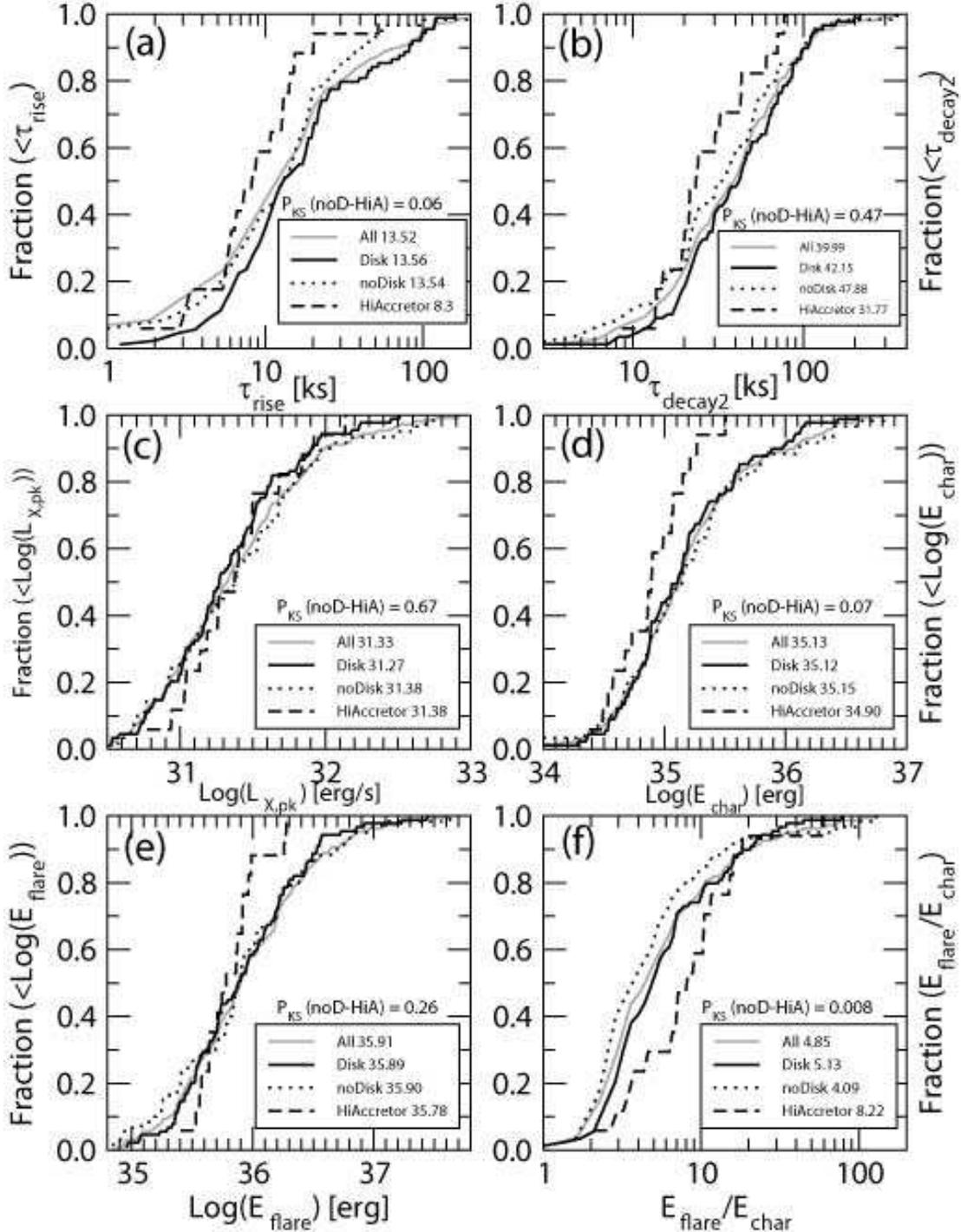}
\caption{\footnotesize Cumulative distributions of flare rise
($a$) and decay ($b$) timescales; X-ray flare peak luminosity
($c$); energy from the ``characteristic'' state within flare
duration time; ($d$); flare energy ($e$); and ratio of energies in
flare and ``characteristic'' states ($f$).  Line types indicate
samples: all 216 flares (grey), 89 flares from MIR disk stars
(solid black), 60 diskfree stars (dotted black), and 17
high-accretor stars (dashed black). Legends indicate median
values, and KS test probability compare diskfree and high-accretor
stars. \label{fig_energy_contrib}}
\end{figure}

Figure \ref{fig_energy_contrib} $a$-$b$ show the distributions of
flare rise and decay times stratified by disk and accretion
properties.  There is a hint that high-accretors have
systematically shorter flare timescales compared to the rest of
the bright COUP flare sample.  The evidence is only suggestive
because our sample of high-accretors is small (17 flares from 12
sources) and it is only marginally significant when measured by a
KS test ($P = 0.06$ when rise times for high-accretors are
compared to diskfree systems).  Figure \ref{fig_energy_contrib}
$c$ shows that the distribution of flare peak luminosities are
indistinguishable for diskfree, MIR disk, and accretion disk
systems.

We evaluate the total energies of each of the 216 flares as the
difference between the time-integrated flare energy $E_{flare}$
and the energy from the non-flare ``characteristic'' state
$E_{char}$ within the duration of the flare, ${\rm t}_{flare2}-
{\rm t}_{flare1}$.  We estimate $E_{flare} \approx L_{X,pk} \times
\tau_{decay2}$ and $E_{char} = L_{X,char} \times ({\rm
t}_{flare2}-{\rm t}_{flare1})$, where X-ray luminosity from the
``characteristic'' state, $L_{X,char}$, was taken from our TRS
analysis (Paper I). $E_{char}$ is systematically lower in flares
from high-accretors (panel $d$);  this is due to the shorter flare
durations (above) and lower $L_{X,char}$ in accreting
systems\footnote{In our flare sample the effect of lower
$L_{X,char}$ in accreting systems is not strong. A KS test gives
only marginally significant difference ($P_{KS} = 0.1$) in
$L_{X,char}$ between high-accretors and diskfree stars. Median
values of $\log (L_{X,char})$ are 30.16 erg~s$^{-1}$  for diskfree
stars, 30.10 erg~s$^{-1}$  for MIR disk stars,  and 30.10
erg~s$^{-1}$ for high-accretors.}. The latter effect is the
well-established suppression of time-integrated X-ray emission in
accreting vs. non-accreting PMS systems \citep[][and references
therein]{Gregory07}. Due to shorter flare timescales, $E_{flare}$
in high-accretors shows a somewhat narrower distribution than that
of other stars but the difference is not statistically significant
(panel $e$).

A more interesting effect is seen in the ratio
$E_{flare}/E_{char}$ which is systematically larger in
high-accretors that other stars. The median value
$E_{flare}/E_{char} \simeq 10$ compared to 5 in diskfree stars
($P_{KS}=0.008$, Figure \ref{fig_energy_contrib}$f$). In the
840~ks of the COUP observation, a typical single bright flare with
a duration of 90~ks (median of ${\rm t}_{flare2}-{\rm t}_{flare1}$
for all 216 flares) may increase the time-integrated source X-ray
luminosity $1.5-2$ times if $E_{flare}/E_{char} = 5-10$. For a
shorter more typical $Chandra$ exposure,  a 50~ks bright flare
($<22\%$ of flares analyzed here have durations $<50$~ks) will
change the time-integrated source X-ray luminosity $3-5.5$ times
if $E_{flare}/E_{char} = 5-10$.

The major result of the section is that no difference is found in
flare duration, peak luminosity and total energy between flares
occurring in disky and diskfree systems. However, as a tentative
finding flares from high-accreting disky stars seem to be somewhat
shorter and thus have weaker total X-ray energies than the rest of
the analyzed flares. This tentative finding is supported by two
statistically significant findings: 1. super-hot COUP flares are
found to be shorter than cooler COUP flares (Paper~I) and 2.
super-hot flares preferentially present in accreting systems (\S
\ref{superhot_section}).

\section{DISKS MAY TRUNCATE PRE-MAIN SEQUENCE MAGNETOSPHERES
\label{truncation_section}}

\begin{figure}
\centering
\includegraphics[angle=0.,width=6.5in]{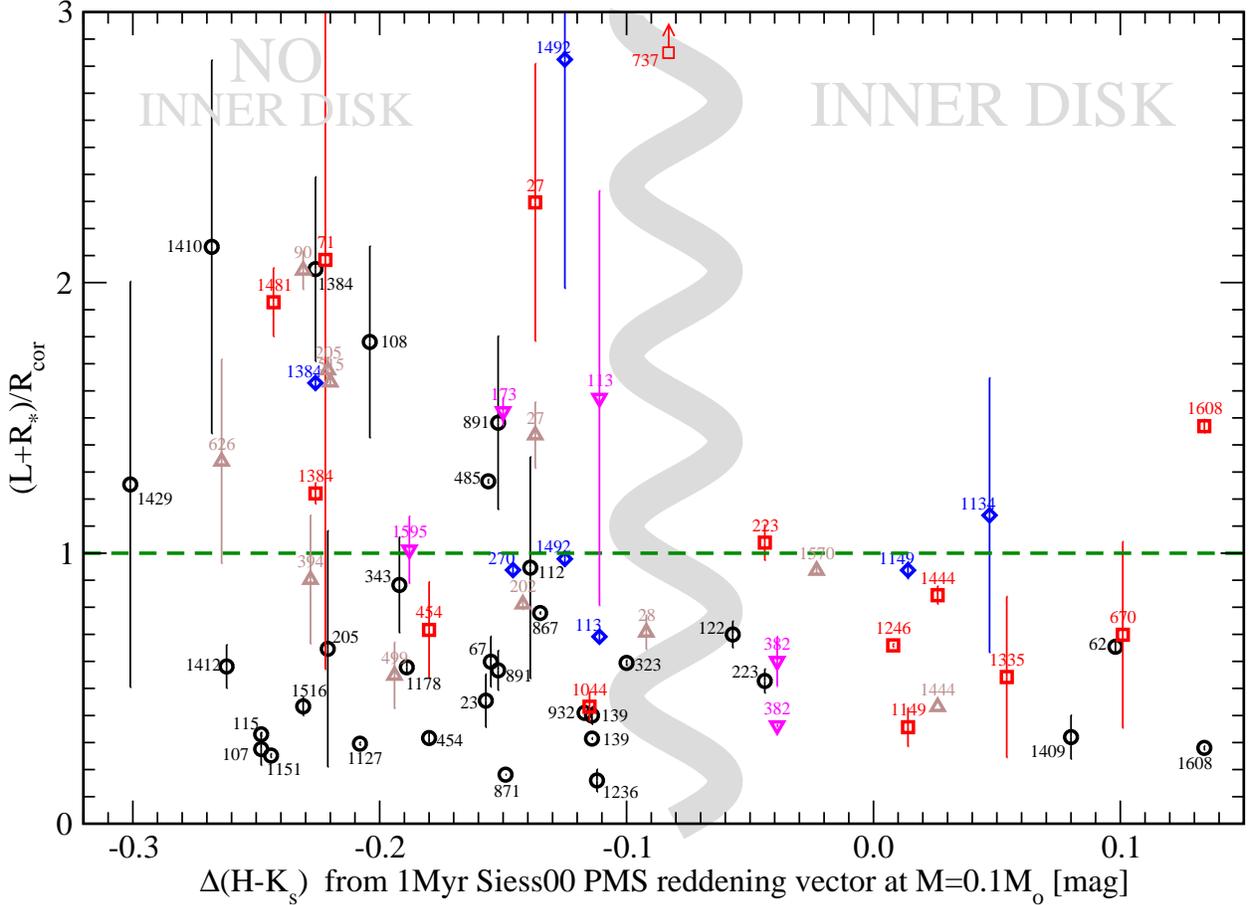}
\caption{Inferred sizes of coronal flaring structures scaled to
stellar corotation radii $vs.$ $K_s$-excess inner disk indicator.
Results are shown for 69 flares from 56 COUP stars with available
data.  Flares are labelled by corresponding COUP source numbers.
Symbols distinguish flare morphologies: 32 ``typical'' flares
(black circles); 14 ``step'' flares (red boxes); 7
``slow-rise-and/or-top-flat'' flares (blue diamonds); 11
``incomplete'' flares (brown triangles); and 5 ``other'' flares
(magenta triangles). The grey serpent line roughly discriminates
between stars with and without inner disks. The vertical bars
represent boundaries of the inferred loop size ranges (Paper~I)
with symbols positioned at the mean of those ranges.
\label{fig_L_vs_dHKs}}
\end{figure}

\begin{figure}
\centering
\includegraphics[angle=0.,width=6.5in]{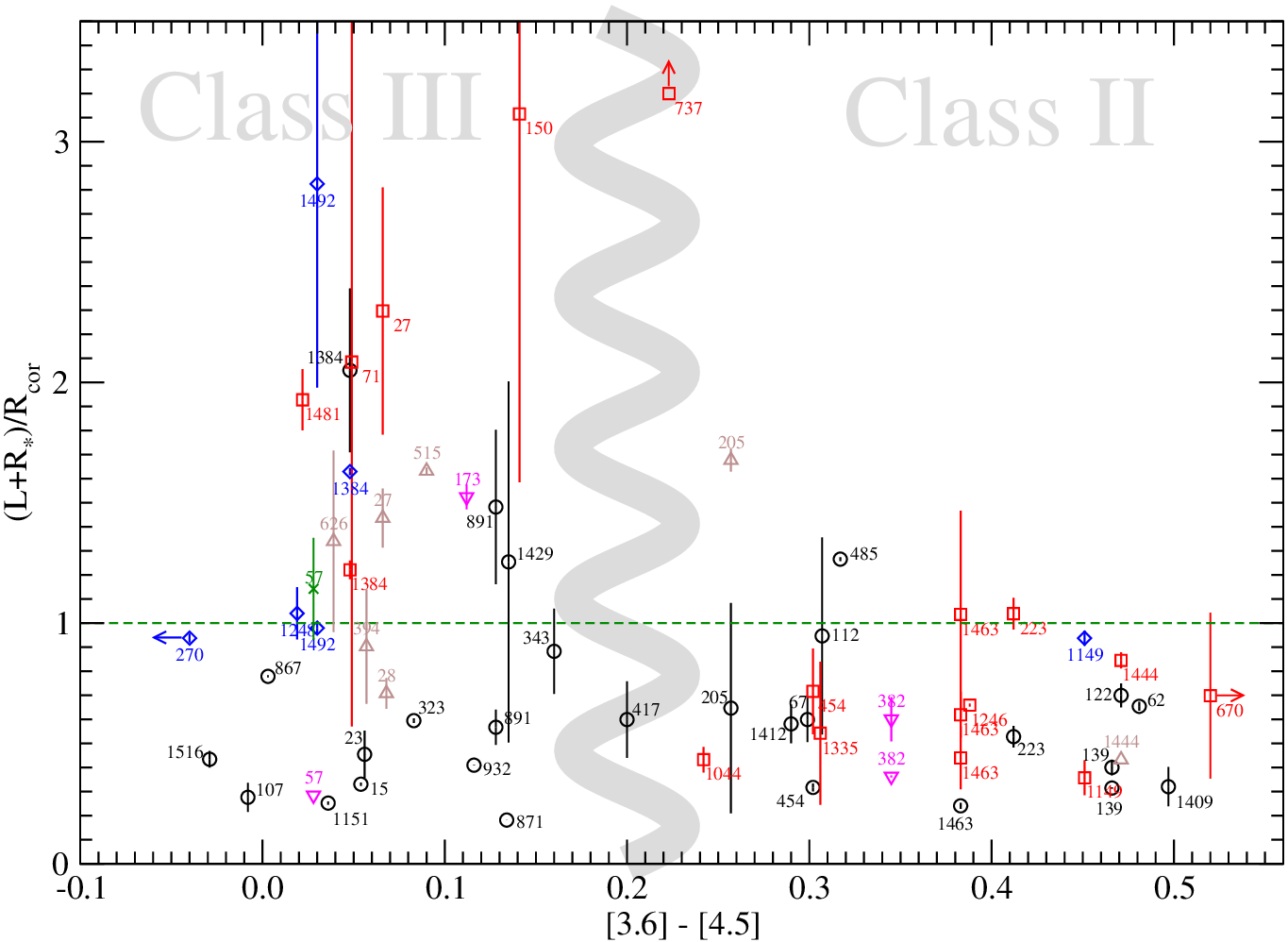}
\caption{Inferred sizes of coronal flaring structures scaled to
stellar corotation radii $vs.$ [3.6] $-$ [4.5] MIR disk indicator.
Results are shown for 63 flares from 47 COUP stars with available
data.  See Figure~\ref{fig_L_vs_dHKs} for symbols.
\label{fig_L_vs_ch12}}
\end{figure}

Figures \ref{fig_L_vs_dHKs},  \ref{fig_L_vs_ch12} and
\ref{fig_L_vs_ewca} compare the flare loop lengths inferred from
analysis of the X-ray spectral evolution of the decay phases
(Paper I) with our disk and accretion indicators. After careful
investigation of various measures of magnetospheric size, we
choose to examine the ratio of the coronal extent of the loop as
measured from the star center, $L+R_{\star}$, to the Keplerian
corotation radius $R_{cor}$ determined by the stellar rotation
rate.  We are thus less interested in the loop size measured in
meters than the relative sizes of the loop to the likely location
of the inner edge of the disk.  This measure reduces variations
associated with star mass, age and rotation and focuses on the
question of the relationship between PMS disks and magnetospheres.

Figure \ref{fig_L_vs_dHKs} plots $(L+R_{\star})/R_{cor}$ against
the NIR $\Delta(H-K_s)$ disk indicator. The vertical lines for
each flare do not represent error bars, but are the low and upper
boundary of the inferred loop size ranges (Paper I) with symbols
positioned at the mean of those ranges. We find that, except for
COUP \#~1608\footnote{This outlier is a visual double unresolved
by 2MASS, and we suspect the NIR photometry is unreliable. It lies
outside the field of the $Spitzer$ IRAC images, so its MIR
properties are unavailable.}, coronal structures responsible for
flares detected from a dozen sources with NIR inner disks do not
exceed $R_{cor}$, while coronal structures responsible for $\sim
40\%$ of flares from 43 sources without NIR inner disks exceed
$R_{cor}$. Some of these flares in diskfree systems arise from
loops reaching $\ga 2 \times R_{cor}$. This pattern is present in
each morphological flare type (typical, step, double, slow
rise/flat top) as indicated by different symbols.

The same pattern is seen when  $(L+R_{\star})/R_{cor}$ is plotted
against the MIR disk indicator $[3.6]-[4.5]$ (Figure
\ref{fig_L_vs_ch12}) and the accretion indicator ${\rm EW(Ca~II)}$
(Figure \ref{fig_L_vs_ewca}).  Virtually all of the flares whose
inferred sizes exceed the host star's corotation radius are
diskfree and non-accreting systems, while disky and accreting
flare loops all lie within the corotation radius.  The MIR plot
adds nine flares from five sources which were not available in the
NIR plot; all of these follow the NIR trend.  Two outliers, COUP
\#205 and 485, are those for which inner part of the disk is
likely cleared of circumstellar material (judging from their NIR
colors). Again, the trend is strong and applicable to each
morphological type of flares. The sample of stars with strong
${\rm EW(Ca~II)}$ emission indicating active accretion is smaller
than those with infrared photometric excesses, but the trend of
smaller loop sizes in accreting systems is still clearly seen.

These three figures provide strong and consistent support for a
model where protoplanetary disks truncate PMS magnetospheres. We
do not know of any selection effect that would have produced this
pattern in a spurious fashion.  The plots in Figure
\ref{fig_corona_extent_suppl} help elucidate this trend. Here the
symbol colors represent the classification of disky (blue) and
diskfree (red) stars based on MIR colors, and only the mean value
of inferred flare loop size is shown. Recall from Figure
\ref{fig_rotation} that, due to the well-established connection
between disks and rotation, corotation radii scaled to stellar
radii are systematically larger for disky compared to diskfree
stars (the difference is roughly a factor of two). This difference
in $R_{cor}/R_{\star}$ is the major contributor to the difference
in $(L + R_{\star})/R_{cor}$ between the disky and diskfree stars
seen in Figures \ref{fig_L_vs_dHKs}-\ref{fig_L_vs_ewca}.   If the
loop sizes $L/R_\star$ were to be considered without normalization
to the corotation radii, the difference between the two classes
becomes marginal\footnote{To avoid observational bias of Class~III
stars towards higher $R_{star}$ and thus restricting the flare
sample to stars with stellar radius in the range of
$R_{\star}=[1-3]$~R$_{\odot}$ gives the following results: KS test
shows no statistical difference ($P_{KS} = 0.2$) in $L/R_{\star}$
between Class~II and Class~III; median values of
$L/R_{\star}=3.2(2.4)$ for Class~II(III) suggest that
$L/R_{\star}$ in Class~II stars is 1.3 times larger than that of
Class~III.}.

We thus find that X-ray coronal extents are somewhat similar in
diskfree and disky systems but, due to the well-established fact
that diskfree stars are faster rotators, the X-ray flares often
exceed the corotation radius in these systems. In contrast, X-ray
flare loops on disky stars never exceed the corotation radius,
although in some cases they reach the corotation radius which is
also the likely truncation radius for the circumstellar disk. The
fact that X-ray loops of disky stars are close to but never exceed
{\bf the one corotation radius} supports long-standing models of
star-disk magnetic interaction at this inner edge involving
accretion, outflow ejection, and regulation of the stellar angular
momentum. The very large loop sizes seen in all types of PMS stars
point to confinement by strong magnetic fields of T-Tauri stars
which, particularly in rapidly rotating diskfree systems, are
capable of withstanding the effects of centrifugal forces
\citep{Jardine99}.

\section{Super-hot flares and disks \label{superhot_section}}

Paper I describes the selection of 73 of the 216 COUP flares as
`super-hot' with peak plasma temperatures $T_{pk} \ga 100$~MK.
Similar flares have been occasionally reported in other PMS
systems such as the diskfree binary PMS star system V773 Tau
observed with ASCA \citep{Tsuboi98}, two embedded young systems in
the NGC~2264 star forming region \citep{Simon05}, and about half
of the COUP flares studied by \citet{Favata05}.  However, here we
have a sufficiently rich sample to study super-hot PMS flares as a
class. This is made possible by the large Orion population, the
unusually long COUP  exposure, and our new highly-sensitive flare
analysis techniques.

There is some concern that the $Chandra$ telescope cannot discern
differences in plasma temperatures above $\sim 100$~MK due to the
rapid decline in mirror reflectivity above $\sim 8$~keV.  However,
we explain in detail in Appendix B of Paper I and the Appendix
below that, when high-signal flares are considered, that
discriminations between 100~MK and $\ga 200$~MK peak temperatures
as well as between $\la 100$~MK and $\ga 100$~MK are possible
using the median energy as a temperature indicator (our MASME
technique described in Paper I). We argue that $\ga 200$~MK
super-hot peak temperatures were missed by the more traditional
time-resolved spectroscopy techniques.

To check on the applicability of the Reale model to unusually large and hot 
loops the detailed time-dependent hydrodynamic simulation was applied to a
typical super-hot flare of the COUP source \#1343 with observed peak flare 
temperature of a few-to-several hundred MK and derived sizes of associated coronal 
flaring structures of $\sim 10^{12}$~cm \citep[section 4.1 in][]{Favata05}. 
The model achieved peak temperature of $\sim 200$~MK and matched both, the observed flare spectrum 
and lightcurve. The loop plasma was heated rapidly to $\sim 200$~MK on time-scale of
1 hour following by explosive chromospheric evaporation with chromospheric 
plasma reaching the loop apex on similar time-scale of 1 hour. This was followed
by the decay phase from the nearly equilibrium state governed by conduction
and radiation cooling processes on a time-scale of a few to several hours.

Table \ref{tbl_correl_new} shows that the flare peak temperatures of stars with
MIR disks are systematically higher than peak temperatures for stars without MIR
disks at high statistical significance ($P_{KS} = 0.9\%$). It is not clear why
the effect is not seen in $K$-band excess systems.  Figure \ref{fig.peakTdisk}$a$
shows that this effect is even more prominent when disk-free stars are compared
to high accreting stars based on the Ca~II emission line indicator for accretion.
Flares from MIR-excess disk stars (solid black distribution) are on average
hotter than those from non-disk stars (dotted) and, among flares from disk stars,
those from high-accretors (dashed) are the hottest. As a consequence,
high-accretors have the largest fraction of super-hot ($T_{obs,pk} > 100$~MK)
flares (53\%) compared to that of stars with MIR disks (40\%)
and diskfree stars (27\%). The effect of systematically hotter flares in disk
and highly accreting stars becomes even more prominent (at a significance level
of $P_{KS} = 0.02\%$) when only flares from $M<2$~M$_{\odot}$ stars are
considered (Figure \ref{fig.peakTdisk}$b$). Even when uncertainties on
individual values of $T_{obs,pk}$ (which can be large for very high
temperatures; Appendix B in Paper~I) are taken into account through Monte-Carlo
simulations, this significance level does not exceed $P_{KS}=1.5\%$\footnote{
We simulated and compared 10000 temperature distributions for each of the sample
(diskless, disk and high-accretors), with individual temperature values randomly
drawn from Gaussian distributions with mean equal to the measured $T_{obs,pk}$ 
and variance as an average error reported in Appendix B of Paper~I. We find 
that $(50\%,68\%,90\%)$ of 10000 resulting significance levels from both, the 
K-S tests between diskless vs. disk and diskless vs. high-accretors comparisons,
do not exceed values of $P_{KS} = (0.3\%,0.5\%,1.5\%)$, respectively.}. We thus
find strong evidence that super-hot flares are preferentially associated with 
PMS stars undergoing substantial accretion.

Disky stars which possess super-hot flares, including accretors,
tend to have spectral types late-K and M corresponding to masses
$M \la 1 - 2$~M$_\odot$, while the majority of diskfree stars with
super-hot flares are more massive with spectral types early-K
through F (thin dashed line in Figure \ref{fig.peakTdisk}$b$). This 
latter group is not large: for sources with known
masses, 66\% of the super-hot flares are produced by stars with
$M<1$~M$_{\odot}$ while only 15\% are produced by stars with
$M>2$~M$_{\odot}$. This can be attributed to the rarity of
intermediate-mass stars compared to lower mass stars in the
Initial Mass Function\footnote{Out of 161 COUP stars analyzed here
83 have known masses of $M<1$~M$_{\odot}$ and 23 have masses of
$M>2$~M$_{\odot}$. Out of 57 stars producing super-hot flares 27
have known masses of $M<1$~M$_{\odot}$ and 7 have masses of
$M>2$~M$_{\odot}$. The fraction of low-mass super-hot stars
($27/83$) is comparable to that of higher-mass stars ($7/23$).}.
The bottom line seems to be that in low-mass stars
($M<1$~M$_{\odot}$) the appearance of super-hot flares is
connected to the presence of active (accreting) disks, however
super-hot flares may also arise in massive ($M>2$~M$_{\odot}$)
diskfree stars.

\section{Magnetic fields geometries \label{magnetic_fields.section}}

Our analysis can provide indirect access to the geometry and
strength of the large-scale magnetic fields responsible for
confining the X-ray emitting plasma during the bright PMS flares
studied here. As described in Paper I, the flare decay model gives
estimates of the flare plasma peak emission measure ($EM_{pk}$),
coronal loop size (we use here the mean value of the size range,
$L$), and plasma temperature ($T_{obs,pk}$). If we add an
assumption concerning the ratio of the cylindrical loop
cross-sectional radius to the loop length $\beta$, we can derive
the plasma electron density $n_e$ from the emission measure and
loop length. The magnetic field confining the plasma can then be
estimated assuming pressure equilibrium,
\begin{equation}
B_{eq} \simeq (8 \pi \times 2 n_e k T^\prime_{pk})^{1/2} ~ {\rm where} ~ \\
n_e \simeq (EM_{pk}/(2 \pi \beta^{2} L^{3}))^{1/2}.
\label{Bfield_eqn}
\end{equation}
\noindent Following past flare models, we adopt a 10:1 cylindrical
geometry, $\beta = 0.1$.  $T^\prime_{pk}$ used here is the plasma
temperature at the loop apex, which is hotter than the observed
X-ray temperature integrated over the entire loop according to
$T^\prime_{pk} = 0.068 \times T_{obs,pk}^{1.2}$ where both
temperatures are in units of K \citep{Reale02,Favata05}.

Figure \ref{fig_ultrahot_5} shows the resulting inferred plasma
densities (panel $a$) and magnetic field strengths (panel $b$) for
all (147 of the 162) flares with known loop sizes and stellar
radii of their host stars. Plasma density $n_e$ is anticorrelated
with the loop size $L$ as expected from equation (1); the grey
line shows the relation $n_e \propto L^{-3/2}$ expected for
constant emission measure and stellar radius.   Estimated plasma
densities\footnote{ We note that the highest loop plasma densities
derived here from modeling flare events are comparable to the high
densities inferred from high-resolution X-ray spectra which are
usually attributed to shocks at the base of accretion streams
\citep[e.g.,][]{Kastner02, Schmitt05, Drake05, Argiroffi07,
Gudel07b, Huenemoerder07}.  This indicates that plasma density
alone may not be a reliable discriminant between flare and
accretional X-rays.} range  from $4 \times 10^9$~cm$^{-3}$ for
very large loop sizes ($L \sim 10$~$R_{\star}$) to $1 \times
10^{12}$~ cm$^{-3}$ for small loop sizes of $L \la
0.2$~$R_{\star}$.

Figure \ref{fig_ultrahot_5}$b$ plots the equilibrium magnetic
field strengths obtained from equation (1) of individual COUP
flares against the inferred flare loop lengths.  But the
relationship between magnetic field strength and loop length is
also a function of field geometry.  The simplest geometry often
assumed for PMS stars is a dipole,
\begin{equation}
B = \frac{B_{ph}}{(L/R_{\star}+1)^3}
\end{equation}
where $B_{ph}$ is the photospheric magnetic field which can be
directly measured. Figure \ref{fig_ultrahot_5}$b$ shows these
relations for several photospheric field strengths in the range
$1-6$~kG consistent with measurements of Zeeman broadening and
circular polarization of PMS photospheric lines
\citep{Johns-Krull99, Symington05, Johns-Krull07, Donati07}.

Figure \ref{fig_ultrahot_5}$b$ shows that for most of the flares
the magnetic fields estimated assuming pressure equilibrium are
consistent with the assumption of dipole geometry with
photospheric magnetic strengths comparable to the observed values
of $1-6$~kG. These include all types of systems: diskfree stars
(red circles), MIR disk stars (blue circles), active accretors
(green boxes), and higher mass stars (black triangles). Thus the
majority of the flares from our flare sample are associated with
the extended, dipole-like loops, and as already discussed in
Paper~I are different from the previously reported smaller stellar
flares which take place in the complex surface field regions
likely making up the ``characteristic'' level of X-ray emission.

However about 40 flares have inferred field strengths too strong
for their inferred loop lengths and/or have loop lengths too long
for any realistic dipole model.  Nearly all of these discrepant
flares have super-hot peak temperatures (magenta circles in Figure
\ref{fig_ultrahot_5}(b)).  If a dipolar topology and pressure
equilibrium holds, the implied surface magnetic field is tens of
kilogauss, far above reasonable values.

This discrepancy can not be fully explained by uncertainties in
inferred peak flare temperatures. Both Appendix B of Paper I and
the Appendix below provide detailed evidence that the existence of
super-hot peak temperatures with $T_{obs,pk} > 100$~MK, and often
$> 200$~MK, is reliable. The uncertainties of these super-hot
temperatures shown in Figure~15 of Paper~I propagate into loop
size uncertainties of 40\% and magnetic strength uncertainties of
10\%. The confidence regions of most super-hot flares in Figure
\ref{fig_ultrahot_5}$b$ thus lie above the locus of dipolar fields
with photospheric strength of $B_{ph} \sim 10$~kG.

As discussed in \S \ref{superhot_section}, super-hot flares appear
to be associated with the presence of a disk and high accretion.
We thus emerge with evidence for anomalous magnetic fields
associated with accreting PMS stars. Super-hot flares typically
arise in large coronal loops $L/R_{\star} \ga 2$ but still inside
the corotation radius. In \S \ref{disc_stardisk.section}, we
speculate that such anomalous fields may result from the
distortion of magnetic topologies or the thickening of magnetic
loops by the process of accretion.

\section{Discussion \label{discussion_section}}

\subsection{Does flaring occur in star-disk magnetic fields?
\label{disc_stardisk.section}}

The first result we encounter in this study is negative: disks
appear to have no effects on flare morphology, timescales, or
energetics (\S \ref{morphology_section}-\ref{energetics_section}).
This is not a trivial finding; there is no reason to believe that
flares arising in field lines attached to the inner rim of the
circumstellar disk would have the same power, plasma properties
and temporal evolution as flares from magnetic loops anchored in
the stellar surface.   Reconnection in star-disk loops can occur
in several ways: by magnetic interactions between star and disk
fields near the corotation radius \citep{Hayashi96}, by stochastic
fluctuations in the accretion rate near the corotation radius
\citep{Shu97}, or by twisting of star-disk loops from stars
slightly out of corotation with their disks \citep{Birk00,
Montmerle00}.  \citet{Isobe03} calculated the hydrodynamical
response of cool plasma to a sudden reconnection event  in the
middle of $1 \times 10^{12}$~cm star-disk loop.  They find that
the shock reaches both the stellar and disk surfaces in
$1-2$~hours with bulk plasma motions sometimes exceeding 1000~km
s$^{-1}$. Temperatures quickly reach $50-100$~MK but the rise in
emission measure sometimes appears slower and less regular than in
typical solar-type flares.  These might be classified as
slow-rise-flat-top or double flares.  Remarkably, Isobe et al.
find that  the entire inner disk gas can be vaporized by powerful
reconnection events.

Thus, although SRTF and double flares conceivably could arise
exclusively in star-disk loops, our results indicate this is not
the case.  Instead, we find that all flare morphological types
occur equally in both diskfree and disky systems with similar
luminosity and duration distributions.  We therefore conclude that
all flare types likely arise in traditional solar-type magnetic
loops where both footprints are rooted in the stellar surface.
This idea is supported by the similarity of COUP flaring
statistics and plasma elemental abundance anomalies to those seen
in older magnetically active stars \citep{Wolk05, Maggio07,
Stelzer07}.   Figure 11 in Paper I shows that step and double
flares are commonly seen in the contemporary Sun, often because of
a triggered reconnection or a reheating event. Slow-rise-flat-top
flares may be similar to other flares with reconnection
sequentially progressing along multiple magnetic arcades. Thus,
all types of COUP flares may arise from solar-type magnetic
morphologies.

However, the solar-type flare analogy to powerful, long-lasting
COUP flares faces a challenge. \citet{Jardine99} has raised the
issue that very large magnetic loops anchored to rapidly rotating
stars may be destroyed when the centrifugal force of their
confined plasma exceeds the weakening magnetic tension in the
outer part of the loop. \citet{Favata05} considered this
instability to be a strong argument in favor of attaching long
loops to the circumstellar disks.  We tentatively conclude from
our findings that, while this centrifugal force might break some
magnetic loops, others survive to produce the observed powerful
flares in diskfree high rotating COUP stars.

There is one flare parameter which is linked to the presence of a
disk: flares on Class~II and high-accretion stars have
systematically hotter peak flare temperatures (\S
\ref{superhot_section} and Table \ref{tbl_correl_new}). This may
point to star-disk reconnection as modeled by \citet{Isobe03}
where temperatures of 100~MK are easily achieved. Additional
calculations should be made to establish the conditions where peak
temperatures of 200~MK or more occur.  However, it is also
possible that these higher temperatures are a byproduct of
systematically higher surface magnetic fields in Class II systems
which would be capable of confining hotter plasmas.  Younger
systems might have stronger fields due to vigorous dynamo
precesses \citep{Browning07}, a primordial field component
\citep{Johns-Krull07}, and/or accretional processes
\citep{Bessolaz08}.  Hotter peak temperatures could be a byproduct
of the magnetosphere compression by disks discussed in \S
\ref{disk_truncation.section}, or atypical coronal loop geometries
with larger aspect ratios than typical of solar coronal loops,
$\beta \gg 0.1$. Rare examples of such ``thick'' stellar coronal
loops are reported in the literature; for example, the AB~Dor~29
Nov 1997 flare has estimated $\beta \sim 1$
\citep{Maggio00,Reale07}.  We conclude that the correlation
between peak flare temperature and disks is an interesting new
result, but does not appear to have a unique explanation and does
not clarify whether or not reconnection occurs in star-disk field
lines.

\subsection{Magnetospheric truncation by a disk \label{disk_truncation.section}}

Our second observational result is positive: the COUP flares
provide the first direct evidence that the magnetic loops of
powerful X-ray PMS flares can exceed the stellar corotation radius
in diskfree stars, but do not exceed the corotation radius in
disky stars (Figures \ref{fig_L_vs_dHKs}-\ref{fig_L_vs_ewca}).
This finding is presented visually in Figure
\ref{fig_corona_extent_sketch} which shows with realistic relative
scaling the largest inferred X-ray coronal extents for Class~II
and Class~III stars.  The high-order multi-polar component of the
magnetic field is expected to dominate at the stellar surface but
fall off rapidly with height, leaving the dipolar component
dominant at distances above a couple of stellar radii
\citep[Figure 15 in][]{Donati07,Donati08}. Figure
\ref{fig_corona_extent_sketch} also shows schematically a
compression and distortion of magnetic field lines by the
accretion disk.  More realistic field configurations with both
accreting field lines and coronal loops are calculated by
\citet{Jardine06, Long07}.

For a T-Tauri star with radius of $\sim 2$~R$_{\odot}$ and mass of
$\sim 0.5$~M$_{\odot}$, the typical large flaring magnetic loops
are roughly the same for the disk classes with $L/R_{\star} \sim
5$.  But, due to the strong rotational acceleration after disks
are gone, the corotation radii shrink so that loops formerly
$(L+R_{\star})/R_{cor} \la 1$ during their disky phase are often
$1 \la (L+R_{\star})/R_{cor} \la 2$ during their diskfree phase.
Table \ref{tbl_correl_known} shows that, as stars evolved from
Class~II to Class~III phases, the average corotation radii
decrease from $R_{cor} \sim 7$~$R_{\star}$ to $R_{cor} \sim
3$~$R_{\star}$ due to a shortening of the average rotation periods
from $P \sim 9$~day to $P \sim 3$~day.

There are three independent supporting lines of evidence
suggesting that long loop structures can be present in
magnetically active stars when disks are not present.
\begin{enumerate}

\item Even the relatively inactive Sun has helmet-like X-ray
streamers that reach up to $\ga 0.5$~R$_{\odot}$ above the
photosphere.  These occur in the subclass of the solar long decay
events (LDEs).  The analogies between solar LDEs and COUP bright
X-ray flares are discussed in Paper I and in
\S~\ref{disc_LDE.section} below.

\item \citet{Mullan06} have carefully compared the loop sizes
inferred from 106 flares studied with the $EUVE$ satellite from 33
magnetically active main sequence stars, including members of RS
CVn binaries and dMe stars.  They find that stars with $B-V<1.4$
(hotter than M0) generally have loop lengths $L/R_\star < 0.5$
while stars with $B-V>1.4$ (cooler than M0) often exhibit loop
lengths $L/R_\star \sim 1.0-1.5$.

\item Radio VLBI studies have shown that magnetospheres likely
extend several stellar radii in very active stars, filling with
trans-relativistic electrons emitting gyrosynchrotron radiation.
This is indicated by VLBI imaging of the RS CVn binary systems UX
Ari HR 1099, and HR 5110 and the dMe stars YY Gem and UV Cet
\citep{Lang94, Alef97, Benz98, Franciosini99, Ransom02, Ransom03}.
While none of these systems showed loops larger than $\sim 2$
times the photospheric radius, recent VLBI mapping of the nearby
diskfree PMS binary system V773 Tau reveals radio structures
extraordinarily far from the component stars.  Their elongated
magnetospheres resembling huge solar helmet streamers extending
$\simeq 20$~R$_\star$ yet are still anchored to the stellar
surface \citep{Massi07}.

\end{enumerate}
As PMS stars lie at the extreme high-luminosity end of the
$10^{10}$ range of correlated X-ray and radio luminosities
\citep[the Benz-G\"udel relation; see Figure 6 in][]{Guedel02}, it
is reasonable that they also have the largest flare loop sizes.
While the typical large coronal structures we find from COUP X-ray
flares are $\sim 5$ times the stellar radius, even larger
structures are found for some cases. These magnetic structures
must extend significantly beyond the corotation surface, implying
that the loop magnetic fields are, at least temporarily, capable
of withstanding both the thermal pressure of the confined hot gas
and the effect of centrifugal forces \citep{Jardine99}.

Our finding that the coronal structures in slower rotating
Class~II stars do not exceed their corotation surface agrees well
with the T-Tauri coronal models of \citet{Jardine06}.  Combining
average COUP levels of X-ray emission with optical band
measurements of multipolar surface magnetic fields and magnetic
circumstellar disks,  they calculate self-consistent 3-dimensional
force-free field configurations.  Their model emerges with a
complicated combination of closed loops confining X-ray emitting
plasma and open field lines available for accreting cool disk
material or releasing a high velocity wind \citep[see
also][]{Lovelace95, Long07}.  In the model of Jardine et al.,
X-ray coronal extents for diskfree PMS  stars reach
$1.5-2.5$~R$_{cor}$, precisely the range we find in the current
work (Figures \ref{fig_L_vs_dHKs}-\ref{fig_L_vs_ewca}). The
centrifugal stripping of the coronal outer parts proposed for
rapidly rotating main-sequence stars, such as VXR 45 with a
rotation period around $0.2$~days \citep{Jardine04}, is apparently
unimportant for relatively slow rotating T-Tauri stars.

We thus find that, while flare properties are mostly similar in
Class~II and Class~III stars (\S \ref{disc_stardisk.section}),
some differences are found. X-ray coronal structures on Class~II
stars are on average somewhat larger and are more likely to
produce super-hot flares.  This suggests the magnetic fields on
accreting Class~II stars may have distorted magnetic topologies as
\citet{Jardine06} predict.  We discuss this further in \S
\ref{disc_superhot.section}.

\subsection{Comparison with solar long decay events
\label{disc_LDE.section}}

It is reasonable to propose that the majority of the bright long
X-ray flares detected from Orion T-Tauri stars are enhanced
analogs of eruptive solar flare events classified as long decay
events \citep[LDEs,][]{Kahler77}. The {\it Skylab}, SMM and {\it
Yohkoh} space observatories show these flares produce X-ray
emitting arches and streamers with altitudes reaching up to
several $\times 10^{10}$~cm \citep[$\ga 0.2$~R$_\odot$,][]
{Svestka95, Farnik96, Svestka97}. The 24 Jan 1992 X-ray streamers,
which reached up to $L \ga 0.5$~ R$_{\odot}$ in altitude
\citep{Hiei94,Hiei97}, are representative of magnetic structures
emerging from solar LDEs (see inset in Figure
\ref{fig_corona_extent_sketch}). The X-ray lightcurves of LDE
flares last from a few hours to a day, similar to the duration of
the COUP flares studied here (see Figure 14 in Paper I for two
examples).

The origin of giant solar X-ray emitting arches and streamers is
not well-understood.  The most widely accepted model is that the
impulsive flare near the solar surface ($L \la 10^{-2}$~R$_\odot$)
blows open the overlying large-scale magnetic field with
subsequent reconnection of magnetic lines through a vertical
current sheet. This model was developed over many years
\citep[e.g.][] {Sturrock66, Kopp76, Forbes96} and is reviewed by
\citet{Priest02}. Peak observed temperatures in giant solar arches
and streamers are typically of several to $\ga 10$~MK with a wide
range of plasma densities.  For example, the 24 Jan 1992 event
showed a relatively low density $\la 10^{8}$~cm$^{-3}$ while the
15 Mar 1993 event showed densities $\ga 10^{10}$~cm$^{-3}$
\citep{Getman99, Getman00}.  Recall, however, that the peak
luminosities and total energies of these solar flares are far
below those we are studying in COUP stars, roughly $E \sim
10^{30-31}$~ergs $vs.$ $E \sim 10^{35-37}$~ergs in the X-ray band
integrated over the flare.

Confining a solar flare plasma with $n_e \la 10^8$~cm$^{-3}$ and
peak temperatures $T \la 10$~MK requires a local field strength
exceeding 1~G. In the solar corona,  the dipole component appears
to become dominant at 2.5~$R_{\odot}$ \citep{Luhmann98} and the
Sun's magnetic dipole field, only $\sim 1$~G at the surface,
likely falls $\la 0.1$~G at this distance. Small-scale multipolar
fields in active regions have surface strengths around 1000~G, but
these quickly decay at large distances. Thus, the solar fields are
too weak to confine X-ray emitting plasma at distances comparable
to or exceeding the solar radius. The large-scale magnetic field
of T-Tauri stars must thus be far stronger than in the Sun, as
argued by \citet{Jardine06} and others (\S
\ref{disk_truncation.section}, \ref{disc_superhot.section}), and
can sustain giant X-ray arches and streamers with sizes
$L/R_{\star} \sim 1-10$. The recent reported discovery of the two
solar-like radio-flaring streamers in the young binary system
V773~Tau~A reaching up the altitudes of $>20$~R$_{\star}$, but
apparently anchored at the surface of its host star
\citep{Massi07}, further strengthens the idea of the strong
large-scale magnetic fields and giant X-ray flaring structures
anchored at the stellar surface of T-Tauri stars.

We recall in this context our tentative finding (limited by small
samples to a low statistical significance) that flares in
accreting systems are systematically shorter than in other systems
(\S \ref{energetics_section}). If this is true, it might be that
high accretion may prevent very long lasting flares.  In order to
have relatively long-lasting flare events, magnetic stresses must
build up over an extended time, storing magnetic energy
accumulated from shearing and twisting of magnetic field lines.  A
similar process occurs in two-ribbon solar flares
\citep{Priest02}, although with much lower total energies. The
magnetic field must be stable in order to store large stresses. It
is possible that the process of accretion destabilizes the field,
preventing the accumulation of magnetic stresses and forcing
shorter flares.

We thus suggest that it is systematically shorter flare durations
due to disrupted magnetic field configurations that may be
responsible for the well-established reduction in time-integrated
X-ray luminosities of accreting PMS stars compared to
non-accreting stars.  This interpretation is different from past
explanations. \citet{Preibisch05b} suggested that X-ray emission
from accretors is suppressed because it cannot arise in magnetic
field lines which are mass-loaded with disk material.
\citet{Jardine06} argued that the outer magnetosphere of accretors
is stripped by interaction with the disk.  \citet{Gregory07}
proposed that soft X-ray emission is attenuated by dense material
in accretion columns. The true cause of the reduced X-ray emission
of Class~II systems is thus still uncertain.

\subsection{Anomalous super-hot flares in accreting stars
\label{disc_superhot.section}}

We described in \S \ref{superhot_section} and in Paper I a
significant subset of COUP flares exhibiting peak temperatures
exceeding 100~MK, some apparently exceeding 200~MK.   These
temperatures are hotter than any solar flare plasma, and hotter
than nearly all reported stellar flares.  The only comparable
event we have identified is the $T>180$~MK plasma temperature in
the 16 Dec 2005 flare in the RS CVn system II Peg using the hard
X-ray detector (designed for gamma ray burst discovery) on board
the $Swift$ satellite \citep{Osten07}.  While the calibration of
$Chandra$ ACIS median energies to plasma temperatures above $\sim
100$~MK is not precise, we argue in Appendix B of Paper I and the
Appendix below that these events are indeed hotter than other
flares in the sample.

While the super-hot flare phenomenon may have more than one cause,
the clearest relation is to active accretion in lower-mass PMS
stars. These are systems where the stellar radius is small and the
stellar magnetosphere appears truncated by the inner disk at the
corotation radius (\S \ref{disk_truncation.section}).  We show in
\S \ref{magnetic_fields.section} that these temperatures are too
hot to be explained by standard dipolar fields rooted in the
stellar surface. We speculate in \S~\ref{disc_stardisk.section}
that these conditions lead to a compression and intensification of
field strength in the outer regions of the loop where the observed
X-ray emission originates. This might lead to more violent
reconnection and successful confinement of higher pressure
plasmas.  This interpretation qualitatively agrees with recent MHD
computations of accretion funnels through PMS magnetospheres where
distortion of initially dipolar field lines are predicted
\citep{Jardine06, Bessolaz08, Romanova08}.  Other explanations
such as reconnection in disk-derived fields
\citep[e.g.,][]{Hayashi96} seem less attractive given the
similarity of superhot and normal temperature flares in other
properties (flare morphology and energetics).

\section{Conclusions \label{conclusion_section}}

We examine here empirical relationships between magnetic
reconnection flares, protoplanetary disks and accretion. The
current work provides detailed flare modeling of a much larger
dataset than previously available: we consider 216 bright X-ray
flares from 161 PMS stars observed during the 13-day COUP exposure
of the Orion Nebula.  Our sample is larger because, as described
in Paper I, we use data analysis techniques that permit modeling
of fainter flares than feasible with traditional parametric
modeling of variable $Chandra$ ACIS spectra.  We thus have
opportunity to uncover more subtle relationships between flaring,
disks and accretion than were previously possible.  The main
results of our study are as follows:

1. Perhaps with the exception of the $Chandra$ study of NGC~2264,
where CTTS are seen to be more variable than WTTS
\citep{Flaccomio06}, past studies using smaller samples have not
found differences in flare statistics or properties as a function
of PMS evolutionary state or mass: accreting Class II systems
appeared to flare like non-accreting Class III systems
\citep[e.g.][]{Stelzer00,  Wolk05, Favata05, Stelzer07}. We
confirm here that Class II and Class III systems produce X-ray
flares with indistinguishable morphologies. Typical fast-rise
slow-decay flares, composite step and double flares, and the
unusual slow-rise flat-top flares are present in both classes. We
have no clear evidence for distinctive flare types emerging from
star-disk magnetic field lines. Accretion variations producing
optical band variations are not associated with X-ray flares
\citep{Stassun06}. Accretion and magnetic flaring thus appear to
be unrelated, even though the theory of PMS accretion involves
magnetic truncation of disks and magnetic funneling of disk
material onto the stellar surface.  The COUP PMS flares are
consistent with solar-type magnetic structures with both
footpoints anchored in the stellar surface.  Many may represent
analogies of a much less powerful class of solar flares known as
long duration events which produce giant coronal arches and X-ray
streamers.

2. A distinct difference is found in the distribution of loop
sizes. We find that X-ray coronal extents in fast rotating
Class~III stars sometimes exceeds the Keplerian corotation radius,
whereas X-ray magnetospheres in Class~II stars appear to be
truncated at the inner edge of accreting protoplanetary disks.
This directly supports theoretical models of magnetically mediated
accretion, magnetic star-disk rotational coupling, and  disk
confinement of PMS magnetospheres.

3. A related, but statistically less secure, result is that flares
from highly accreting Class~II stars have somewhat shorter
durations and weaker total X-ray energies than Class~III flares.
This might reflect the destabilization of magnetic arcades in
accreting systems, and could account for the reduction of
time-averaged X-ray luminosities in Class II compared to Class III
populations noted in previous studies.

4. A subclass of super-hot flares with peak plasma temperatures
greater than 100~MK is noted for the first time.  These are
inconsistent with formation in normal dipolar magnetic loops
attached to the stellar surface, and appear preferentially in
accreting Class~II systems.  They may reflect compression and
distortion of the large scale magnetospheric topology by star-disk
magnetic interactions, as predicted by recent theoretical
calculations.

\acknowledgements  We thank the anonymous referee for his time and 
useful comments that improved this work. The work was supported by the $Chandra$ ACIS
Team (G. Garmire, PI) through the SAO grant SV4-74018. G.M.
acknowledges contribution from contract ASI-INAF I/088/06/0. This
publication makes use of data products from the Two Micron All Sky
Survey (a joint project of the University of Massachusetts and the
Infrared Processing and Analysis Center/California Institute of
Technology, funded by NASA and NSF), and archival data obtained with
the {\it Spitzer Space Telescope} (operated by the Jet Propulsion
Laboratory, California Institute of Technology under a contract
with NASA).

\appendix

\section{SUPER-HOT FLARES  \label{appendix_section}}

It is difficult to accurately determine temperatures of thermal
components $\ga 100$~MK from $Chandra$-ACIS spectra. To insure
that the peak temperatures of $T_{obs,pk}>100$~MK do indeed
describe the hottest flare plasmas observed in our sample, we
describe here checks that have been performed beyond the analysis
of the Appendix B of Paper I.

For each of the 216 COUP flares treated here, a spectrum has been
extracted within the total time range $[{\rm t}_{flare1}-{\rm
t}_{flare2}]$. Spectra were fitted with WABS~$\times$~MEKAL model
\citep[for compatibility with][]{Getman05} with 0.3 times solar
elemental abundances allowing both temperature and column density
to be free parameters. The following confirmatory results are
obtained. First, for the majority of the flares, the X-ray column densities
obtained from the integrated flare spectral fits are in excellent
agreement with the source column densities obtained from the full 
COUP observation by \citet{Getman05}
that we used as fixed parameters during our flare analysis
(see also Figure~16 in Paper~I). Second, out of the 73 flares with peak flare
temperatures $T_{obs,pk}>100$~MK, only 8 (11\%) have their
integrated flare spectral fits $T_{flare}<40$~MK, while 80\%  of
the flares with $T_{obs,pk}<100$~MK have $T_{flare}<40$~MK (Figure
\ref{fig_super_hot_analysis1}). This confirms a distinct
difference between super-hot and ordinary flares.  Third, high
accreting stars\footnote{The twelve highly accreting starswith
$EW(CaII) < -2$~\AA\ are COUP  \#11, 66, 141, 567, 579, 1044,
1045, 1080, 1096, 1335, 1409, and 1444.} have the largest fraction
of reported super-hot flares (9 out of 17 flares or 53\%), three
more flares from high accretors  have their
$80<T_{obs,pk}<100$~MK. This confirms the discussion in \S
\ref{superhot_section}  that high accretors preferentially exhibit
super-hot flares.

Figure \ref{fig_ultrahot_4} further illustrates spectral
differences between super-hot ($T_{obs,pk}>100$~MK) and ordinary
flares.  To avoid possible confounding effects of differing
absorptions, flare histories and disk properties, we extract the
spectrum within a brief 6~ks interval around the flare peak for
all MIR disk sources in the narrow interval
$21.5<\log(N_H)<22$~cm$^{-2}$. The figure compares combined flare
peak spectra for sources with ordinary (panel $a$) and super-hot
flares (panel $b$). Merging these spectra to improve the
statistical accuracy, the best-fit model parameters of the
composite 6~ks spectra are $N_H = 4.8_{-0.5}^{+0.5} \times
10^{21}$~cm$^{-2}$ and  $kT = 5.5_{-0.8}^{+1.0}$~keV for the
ordinary flares, and $N_H = 6.6_{-0.7}^{+0.6} \times
10^{21}$~cm$^{-2}$ and $kT = 15.2_{-3.6}^{+12.1}$~keV for the
super-hot flares. The composite spectrum of super-hot flares can
not be described by ordinary flare models (compare black and green
lines in Figure~\ref{fig_ultrahot_4}$b$). This again demonstrates
that super-hot flares are the hottest flares of our sample.

\clearpage

\begin{deluxetable}{cccccccc}
\centering \rotate \tabletypesize{\small} \tablewidth{0pt}
\tablecolumns{8}

\tablecaption{Summary of Stellar Properties
\label{tbl_correl_known}}

\tablehead{

\colhead{Quantity} & \colhead{N} & \colhead{Flag} & \colhead{Min}
& \colhead{Max} &
\colhead{Median} & \colhead{Mean$\pm$SD} & \colhead{$P_{KS}$}\\

(1)&(2)&(3)&(4)&(5)&(6)&(7)&(8)}

\startdata
X-ray Absorbing Column Density  (cm$^{-2}$) & & & & & & & \nodata\\
$\log(NH)$ (all available) & 161 & s & 20.0 & 22.8 & 21.6 & $21.5\pm0.6$ & \nodata\\
$\log(NH_{NIRdisk})$ (NIR inner disk) & 56 & s & 20.0 & 22.6 & 21.7 & $21.6\pm0.6$ & \nodata\\
$\log(NH_{noNIRdisk})$ (no NIR inner disk) & 84 & s & 20.0 & 22.3 & 21.5 & $21.4\pm0.6$ & 0.09\\
$\log(NH_{MIRdisk})$ (MIR disk) & 62 & s & 20.0 & 22.8 & 21.8 & $21.7\pm0.6$ &\nodata \\
$\log(NH_{noMIRdisk})$ (no MIR disk) & 50 & s & 20.0 & 22.5 & 21.3 & $21.3\pm0.6$ & 0.02\\
\cline{1-8}
Stellar Mass (M$_{\odot}$) & & & & & & & \\
$M$ (all available) & 128 & s & 0.2 & 5.0 & 0.7 & $1.0\pm0.9$ & \nodata\\
$M_{NIRdisk}$ (NIR inner disk) & 42 & s & 0.2 & 5.0 & 0.7 & $1.1\pm1.1$ & \nodata\\
$M_{noNIRdisk}$ (no NIR inner disk) & 72 & s & 0.2 & 2.3 & 0.6 & $1.0\pm0.8$ & 0.1\\
$M_{MIRdisk}$ (MIR disk) & 44 & s & 0.2 & 5.0 & 0.7 & $1.1\pm0.9$ &\nodata\\
$M_{noMIRdisk}$ (no MIR disk) & 42 & s & 0.2 & 2.3 & 0.6 & $1.0\pm0.9$ & 0.4\\
\cline{1-8}
Stellar Radius ($10^{10}$~cm) & & & & & & & \\
$R_{\star}$ (all available) & 130 & s & 4 & 90 & 18 & $20\pm12$ & \nodata\\
$R_{\star,NIRdisk}$ (NIR inner disk) & 44 & s & 5 & 90 & 16 & $19\pm15$ & \nodata\\
$R_{\star,noNIRdisk}$ (no NIR inner disk) & 72 & s & 4 & 58 & 18 & $21\pm11$ & 0.5\\
$R_{\star,MIRdisk}$ (MIR disk) & 45 & s & 4 & 32 & 14 & $17\pm7$ & \nodata\\
$R_{\star,noMIRdisk}$ (no MIR disk) & 43 & s & 10 & 58 & 17 & $22\pm12$ & 0.3\\
\cline{1-8}
Stellar Rotational Period (day) & & & & & & & \\
$P$ (all available) & 79 & s & 1.1 & 19.5 & 6.5 & $6.7\pm3.9$ & \nodata\\
$P_{NIRdisk}$ (NIR inner disk) & 20 & s & 1.4 & 14.4 & 8.5 & $7.7\pm2.8$ & \nodata\\
$P_{noNIRdisk}$ (no NIR inner disk) & 48 & s & 1.1 & 19.5 & 6.3 & $6.3\pm4.2$ & 0.06\\
$P_{MIRdisk}$ (MIR disk) & 24 & s & 5.2 & 14.4 & 9.0 & $8.8\pm1.9$ & \nodata\\
$P_{noMIRdisk}$ (no MIR disk) & 35 & s & 1.1 & 19.5 & 3.5 & $5.5\pm4.6$ & $<$0.0001\\
\cline{1-8}
Keplerian Co-rotation Radius ($10^{10}$~cm) & & & & & & & \\
$R_{cor}$ (all available) & 67 & s & 22 & 270 & 85 & $90\pm47$ & \nodata\\
$R_{cor,NIRdisk}$ (NIR inner disk) & 14 & s & 39 & 180 & 120 & $107\pm39$ & \nodata\\
$R_{cor,noNIRdisk}$ (no NIR inner disk) & 44 & s & 22 & 270 & 84 & $87\pm50$ & 0.03\\
$R_{cor,MIRdisk}$ (MIR disk) & 21 & s & 54 & 180 & 98 & $108\pm33$ & \nodata\\
$R_{cor,noMIRdisk}$ (no MIR disk) & 31 & s & 22 & 270 & 76 & $83\pm55$ & 0.007\\
\cline{1-8}
Ratio of Co-rotation Radius to Stellar Radius & & & & & & & \\
$R_{cor}/R_{\star}$ (all available) & 67 & s & 1.4 & 20.9 & 5.1 & $5.7\pm3.9$ & \nodata\\
$(R_{cor}/R_{\star})_{NIRdisk}$ (NIR inner disk) & 14 & s & 3.0 & 16.3 & 7.4 & $8.1\pm3.6$ &\nodata \\
$(R_{cor}/R_{\star})_{noNIRdisk}$ (no NIR inner disk) & 44 & s & 1.5 & 20.9 & 4.2 & $5.3\pm4.0$ & 0.01\\
$(R_{cor}/R_{\star})_{MIRdisk}$ (MIR disk) & 21 & s & 3.0 & 17.2 & 7.2 & $8.1\pm3.6$ & \nodata\\
$(R_{cor}/R_{\star})_{noMIRdisk}$ (no MIR disk) & 31 & s & 1.4 & 13.6 & 3.2 & $4.0\pm2.7$ & $<$0.0001\\

\enddata

\tablecomments{Columns 1: Quantity name. Column 2: Number of
flares (Flag$=$f, see Column 3) or sources (Flag$=$s) in a sample.
Column 3: Flag indicates if this is a flare (f) or source (s)
sample. Columns 4-6: Min, max, median values for a considered
quantity's distribution. Column 7: Mean and standard deviation for
a considered quantity's distribution. Column 8: K-S probability
for the assumption that distributions of a considered quantity in
disky and diskfree stars are drawn from the same underlying
distribution.}

\end{deluxetable}

\clearpage

\begin{deluxetable}{cccccccc}
\centering \rotate \tabletypesize{\small} \tablewidth{0pt}
\tablecolumns{8}

\tablecaption{Summary of Flare Properties \label{tbl_correl_new}}

\tablehead{

\colhead{Quantity} & \colhead{N} & \colhead{Flag} & \colhead{Min}
& \colhead{Max} &
\colhead{Median} & \colhead{Mean$\pm$SD} & \colhead{$P_{KS}$}\\

(1)&(2)&(3)&(4)&(5)&(6)&(7)&(8)}

\startdata

\cline{1-8}
Peak Flare Luminosity (erg s$^{-1}$) & & & & & & & \\
$\log(L_{X,pk})$ (all available)  & 216 & f & 30.45 & 32.92 & 31.33 & $31.38\pm0.51$ & \nodata\\
$\log(L_{X,pk,NIRdisk})$ (NIR inner disk)  & 78 & f & 30.44 & 32.51 & 31.35 & $31.36\pm0.46$ & \nodata\\
$\log(L_{X,pk,noNIRdisk})$ (no NIR inner disk)  & 106 & f & 30.53 & 32.92 & 31.38 & $31.41\pm0.52$ & 0.9\\
$\log(L_{X,pk,MIRdisk})$ (MIR disk)  & 89 & f & 30.49 & 32.51 & 31.27 & $31.32\pm0.44$ & \nodata\\
$\log(L_{X,pk,noMIRdisk})$ (no MIR disk)  & 60 & f & 30.57 & 32.77 & 31.38 & $31.40\pm0.53$ & 0.4\\
\cline{1-8}
Peak Flare Temperature (MK) & & & & & & & \\
$T_{pk}$ (all available)  & 216 & f & 19 & 700 & 63 & $141\pm178$ & \nodata\\
$T_{pk,NIRdisk}$ (NIR inner disk)  & 78 & f & 23 & 700 & 73 & $170\pm209$ & \nodata\\
$T_{pk,noNIRdisk}$ (no NIR inner disk)  & 106 & f & 19 & 700 & 59 & $123\pm153$ & 0.5\\
$T_{pk,MIRdisk}$ (MIR disk)  & 89 & f & 19 & 700 & 84 & $181\pm209$ & \nodata\\
$T_{pk,noMIRdisk}$ (no MIR disk)  & 60 & f & 20 & 600 & 57 & $102\pm114$ & 0.009\\
\cline{1-8}
Inferred Loop Size ($10^{10}$~cm) & & & & & & & \\
$L$ (all available)  & 175 & f & 0.4 & 510 & 43 & $64\pm70$ & \nodata\\
$L_{NIRdisk}$ (NIR inner disk)  & 59 & f & 4 & 420 & 43 & $60\pm63$ & \nodata\\
$L_{noNIRdisk}$ (no NIR inner disk)  & 89 & f & 6 & 510 & 43 & $67\pm71$ & 0.8\\
$L_{MIRdisk}$ (MIR disk)  & 68 & f & 4 & 510 & 56 & $79\pm86$ & \nodata\\
$L_{noMIRdisk}$ (no MIR disk)  & 54 & f & 0.4 & 300 & 37 & $54\pm55$ & 0.12\\
\cline{1-8}
Ratio of Loop Size to Stellar Radius& & & & & & & \\
$L/R_{\star}$ (all available) & 147 & f & 0.03 & 45 & 2.4 & $4.4\pm7.3$ & \nodata\\
$(L/R_{\star})_{NIRdisk}$ (NIR inner disk)  & 48 & f & 0.14 & 33 & 2.5 & $4.0\pm5.1$ & \nodata\\
$(L/R_{\star})_{noNIRdisk}$ (no NIR inner disk)  & 81 & f & 0.2 & 46 & 2.4 & $4.6\pm7.9$ & 0.5\\
$(L/R_{\star})_{MIRdisk}$ (MIR disk)  & 52 & f & 0.4 & 46 & 3.0 & $5.9\pm8.2$ & \nodata\\
$(L/R_{\star})_{noMIRdisk}$ (no MIR disk)  & 47 & f & 0.03 & 29 & 1.8 & $2.9\pm4.4$ & 0.009\\
\cline{1-8}
Ratio of Loop Size to Co-rotation Radius & & & & & & & \\
$L/R_{cor}$ (all available) & 81 & f & 0.007 & 5.4 & 0.51 & $0.73\pm0.76$ & \nodata\\
$(L/R_{cor})_{NIRdisk}$ (NIR inner disk) & 18 & f & 0.08 & 1.3 & 0.51 & $0.54\pm0.30$ & \nodata\\
$(L/R_{cor})_{noNIRdisk}$ (no NIR inner disk) & 51 & f & 0.05 & 5.4 & 0.51 & $0.82\pm0.88$ & 0.15\\
$(L/R_{cor})_{MIRdisk}$ (MIR disk) & 30 & f & 0.07 & 5.4 & 0.45 & $0.66\pm0.94$ & \nodata\\
$(L/R_{cor})_{noMIRdisk}$ (no MIR disk) & 33 & f & 0.007 & 2.4 & 0.69 & $0.78\pm0.66$ & 0.19\\
\cline{1-8}
Ratio of Loop Size plus Stellar Radius & & & & & & & \\
to Co-rotation Radius &&&&&&& \\
$(L+R_{\star})/R_{cor}$ (all available) & 81 & f & 0.16 & 5.5 & 0.72 & $0.98\pm0.80$ & \nodata\\
$((L+R_{\star})/R_{cor})_{NIRdisk}$ (NIR inner disk) & 18 & f & 0.28 & 1.5 & 0.66 & $0.69\pm0.32$ &\nodata\\
$((L+R_{\star})/R_{cor})_{noNIRdisk}$ (no NIR inner disk) & 51 & f & 0.16 & 5.5 & 0.88 & $1.11\pm0.90$ & 0.08\\
$((L+R_{\star})/R_{cor})_{MIRdisk}$ (MIR disk) & 30 & f & 0.24 & 5.49 & 0.62 & $0.81\pm0.94$ & \nodata\\
$((L+R_{\star})/R_{cor})_{noMIRdisk}$ (no MIR disk) & 33 & f & 0.18 & 3.11 & 0.98 & $1.14\pm0.75$ & 0.01\\
\cline{1-8}

\enddata

\tablecomments{Columns 1: Quantity name. In case of loop sizes,
summary is provided for a distribution of a mean value of the
inferred loop size ranges. Column 2: Number of flares (Flag$=$f,
see Column 3) or sources (Flag$=$s) in a sample. Column 3: Flag
indicates if this is a flare (f) or source (s) sample. Columns
4-6: Min, max, median values for a considered quantity's
distribution. Column 7: Mean and standard deviation for a
considered quantity's distribution. Column 8: K-S probability for
the assumption that distributions of a considered quantity in
disky and diskfree stars are drawn from the same underlying
distribution.}

\end{deluxetable}

\clearpage

\clearpage

\clearpage

\clearpage

\clearpage

\clearpage

\clearpage

\clearpage

\begin{figure}
\centering
\includegraphics[angle=0.,width=6.5in]{f8.eps}
\caption{Inferred sizes of coronal flaring structures scaled to
stellar corotation radii $vs.$ 8542~\AA\ Ca~II line accretion
indicator. Results are shown for 66 flares from 53 COUP stars with
available information data.  The grey serpent line roughly
discriminates between stars with and without accretion. See
Figure~ \ref{fig_L_vs_dHKs} for symbols.  \label{fig_L_vs_ewca}}
\end{figure}

\clearpage

\begin{figure}
\centering
\includegraphics[angle=0.,width=6.0in]{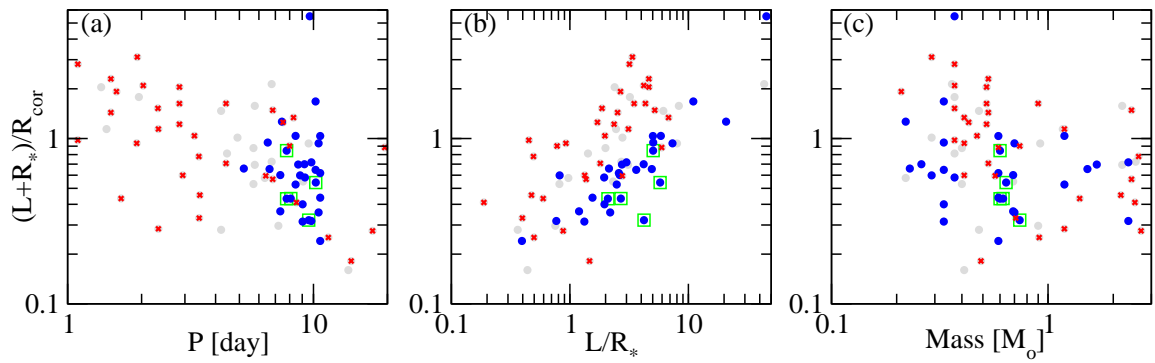}
\caption{Coronal extents of analyzed flaring structures $vs.$: (a)
stellar rotational period; (b) loop size normalized to stellar
radius; and (c) stellar mass.  See Figure~\ref{fig_rotation} for
symbols. \label{fig_corona_extent_suppl}}
\end{figure}

\clearpage

\begin{figure}
\centering
\includegraphics[angle=0.,width=7.0in]{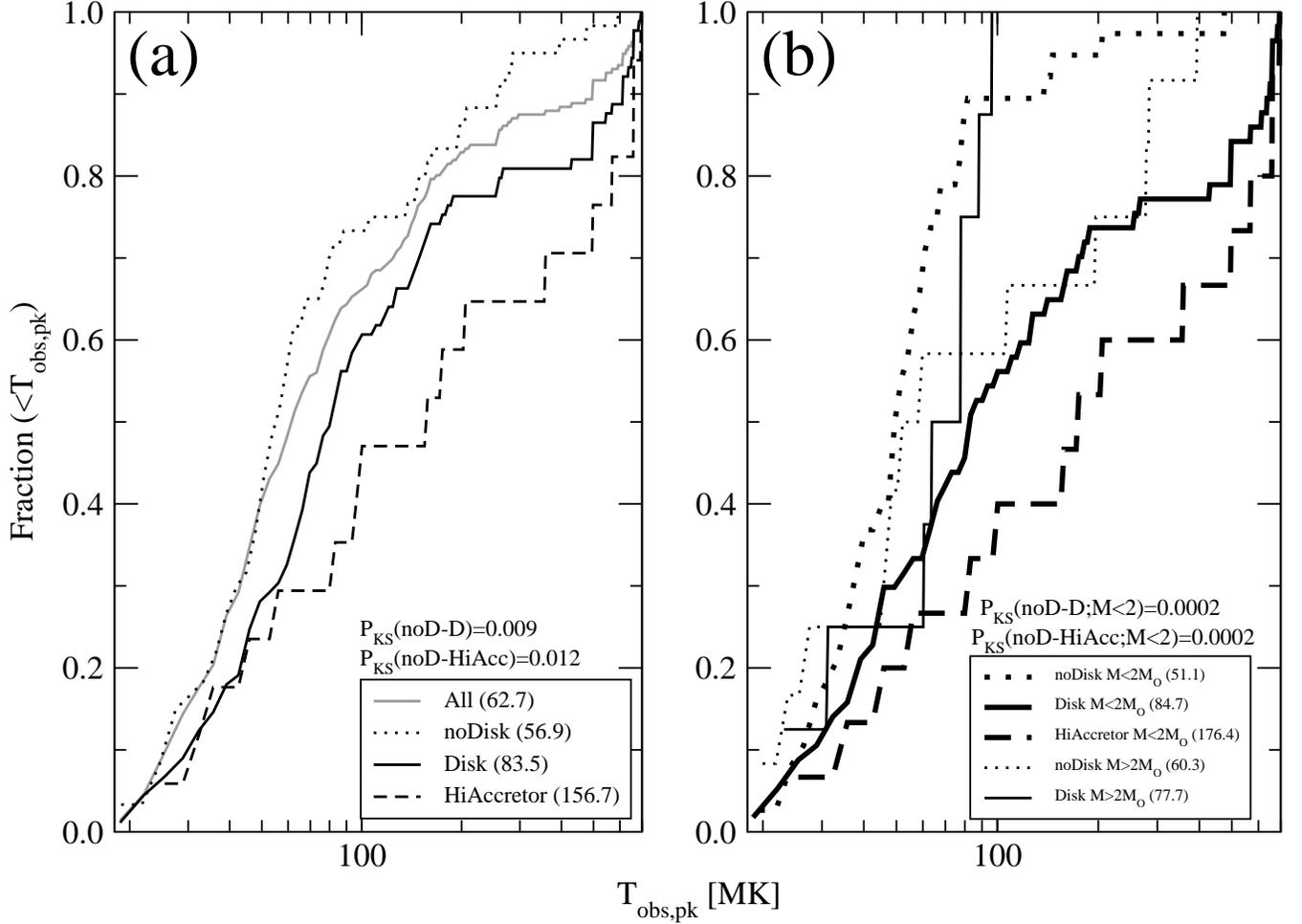}
\caption{(a) Cumulative distributions of flare peak temperature for
all 216 flares (grey line), 89 flares from MIR disk stars (solid black), 60
from diskfree stars (dotted), and 17 from high-accretors (dashed). (b) Mass-stratified cumulative
distributions of flare peak temperature for 57 flares from MIR disk stars with
$M<2$~M$_{\odot}$ (thick solid), 38 from diskfree stars with $M<2$~M$_{\odot}$
(thick dotted), 15 from high-accretors with $M<2$~M$_{\odot}$ (thick dashed),
8 from MIR disk stars with $M>2$~M$_{\odot}$ (thin solid), and 12 from diskfree
stars with $M>2$~M$_{\odot}$ (thin dotted). Legends indicate median values. KS 
test probabilities compare temperature distributions between non-disk and disk 
or high-accretor stars.
\label{fig.peakTdisk}}
\end{figure}

\clearpage

\begin{figure}
\centering
\includegraphics[angle=0.,width=6.5in]{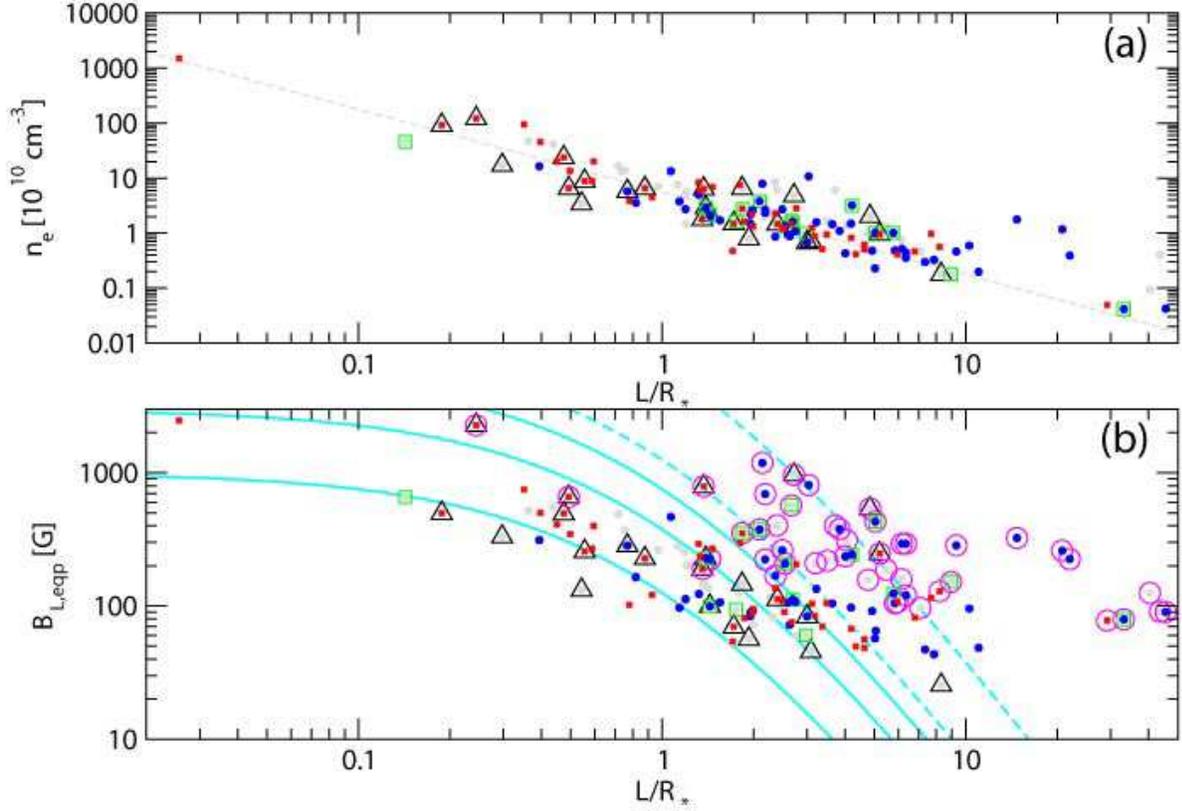}
\caption{Estimated plasma density at flare peak (panel $a$) and
magnetic field from pressure equipartition (panel $b$) are plotted
against inferred loop sizes scaled to stellar radii. Symbols
discriminate subsamples: MIR disk stars (blue circles); no MIR
disk stars (red crosses); uncertain MIR disk stars (grey circles);
highly accreting stars with $EW(CaII) <-2$~\AA\ (green boxes);
masses $M>2$~M$_{\odot}$ (black triangles); and super-hot flares
with $T_{obs,pk}>100$~MK (magenta circles in panel $b$). In panel
$a$, the dashed grey line indicates $n_e \propto L^{-3/2}$. In
panel $b$, the cyan curves represent loci of photospheric dipolar
magnetic fields $B_{ph}=(1, 3, 6)$~kG (solid) and $(10, 50)$~kG
(dashed). \label{fig_ultrahot_5}}
\end{figure}

\clearpage

\begin{figure}
\centering
\includegraphics[angle=0.,width=6.5in]{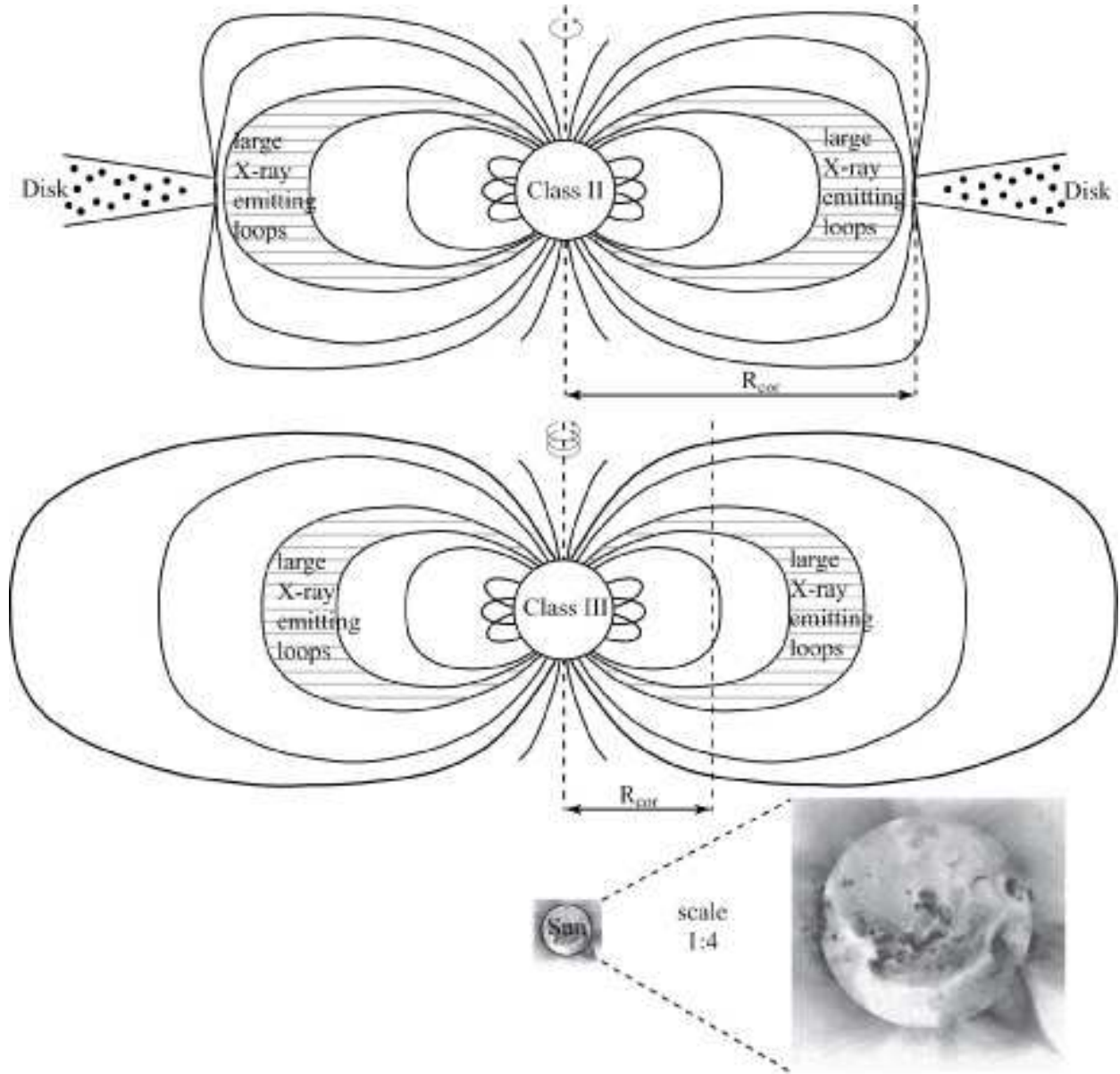}
\caption{Sketch showing typical sizes of the largest X-ray
emitting magnetic structures assuming a dipolar geometry (hatched)
for Class~II and Class~III T-Tauri stars, respectively.   Relative
sizes are shown to scale for typical stars.   The inset shows the
X-ray image of the Sun, both to scale and expanded by a factor of
4 for clarity, showing a large helmet-like X-ray streamers  (24
Jan 1992 flare, $Yohkoh$~SXT).   \label{fig_corona_extent_sketch}}
\end{figure}

\clearpage

\begin{figure}
\centering
\includegraphics[angle=0.,width=6.5in]{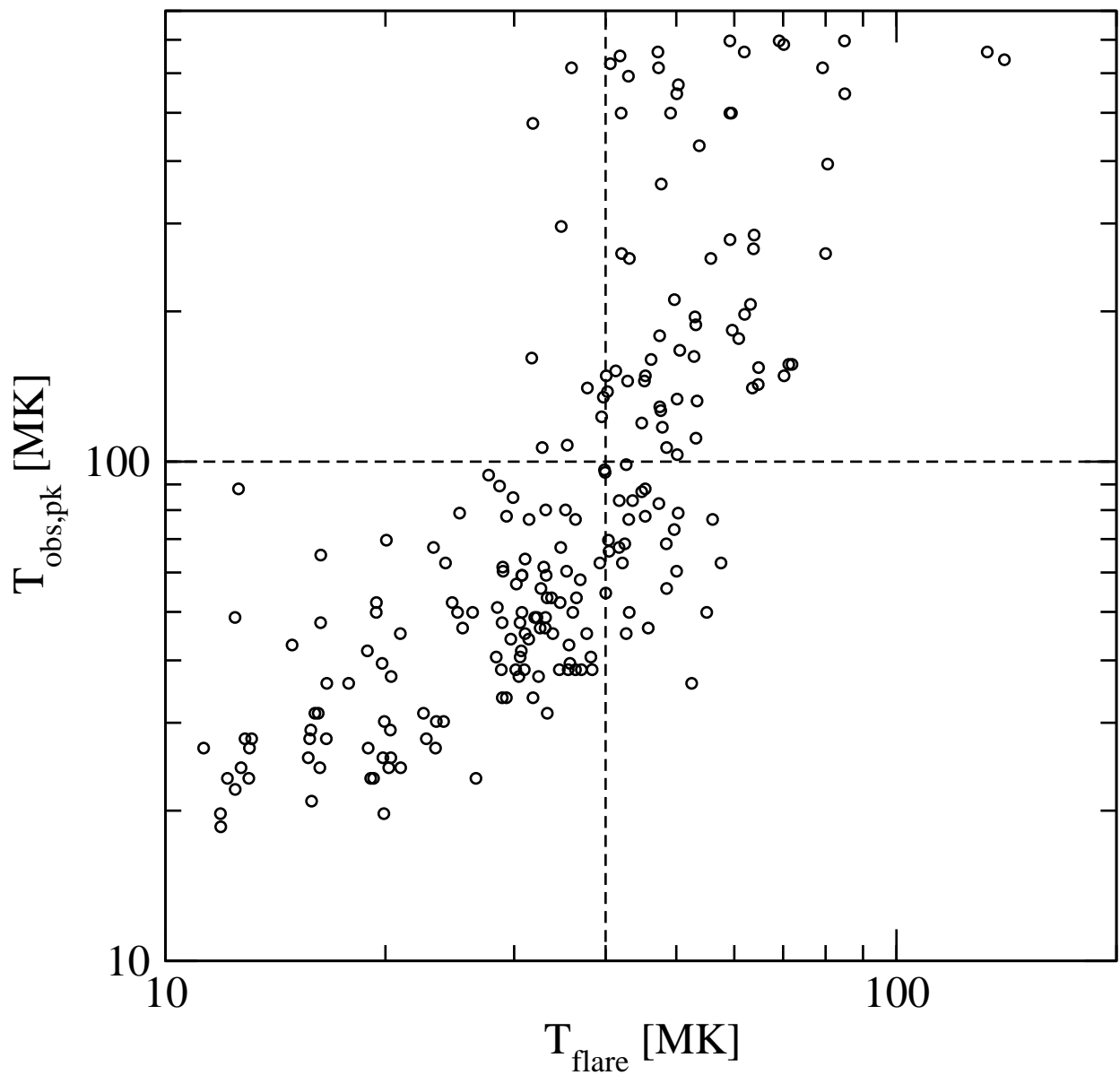}
\caption{For all 216 COUP flares, comparison of the flare peak
temperature from MASME analysis (Paper~I) with the temperature
derived from the spectral fit of the integrated flare spectrum.
Dashed lines show 100~MK and 40~MK, respectively.
\label{fig_super_hot_analysis1}}
\end{figure}

\clearpage

\begin{figure}
\centering
\includegraphics[angle=0.,width=5.5in]{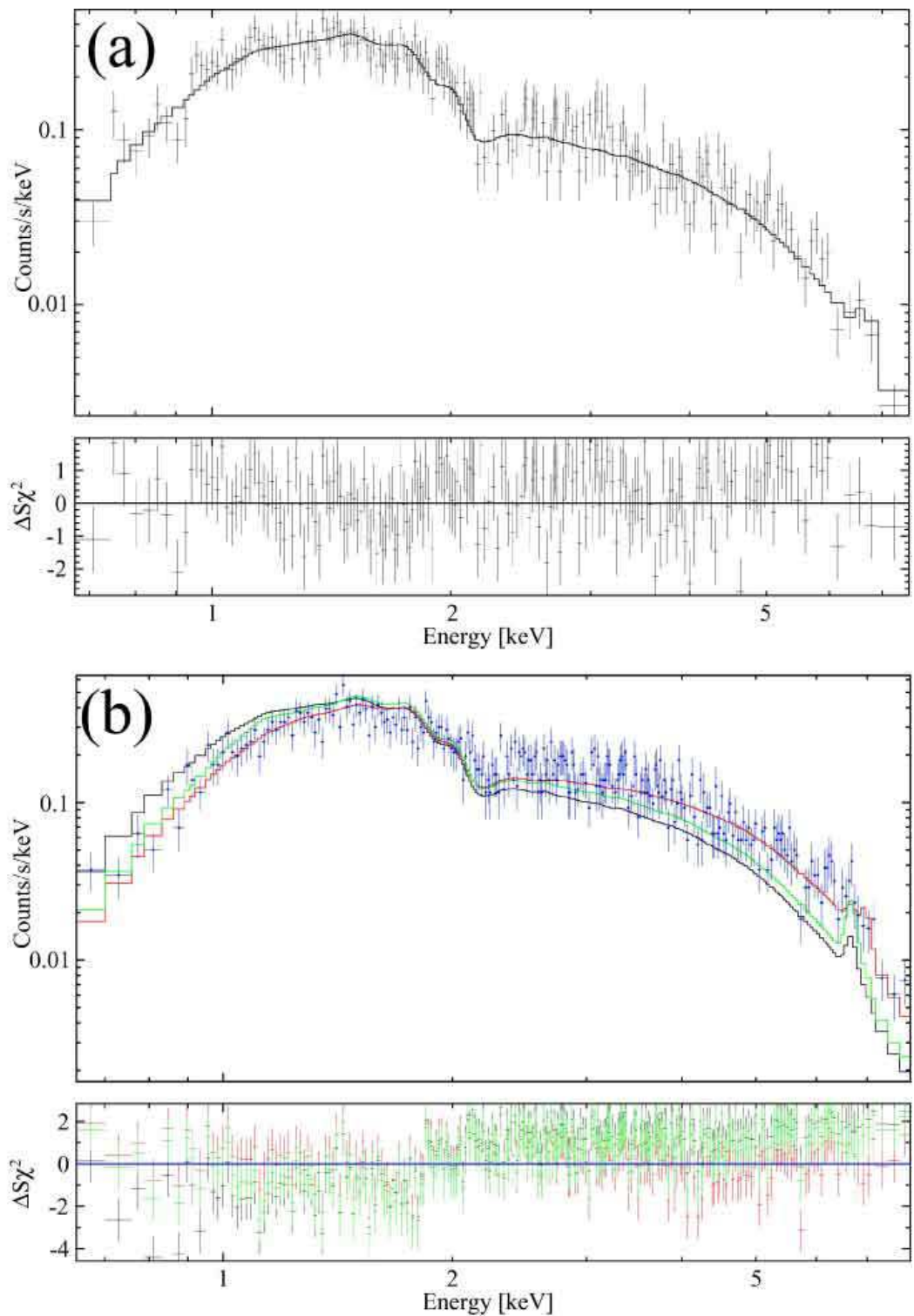}
\caption{\footnotesize Composite flare peak spectra from MIR disk
stars with source X-ray column densities in the narrow range of
absorption.  See Appendix for sample details. (a) Composite of 14
ordinary flares from 12 stars with $T_{obs,pk} < 100$~MK. (b)
Composite of 21 super-hot flares from 14 stars with $T_{obs,pk}
> 100$~MK.  Three spectral models and residuals are:
ordinary flare model (black), super-hot flare model (red),
ordinary flare model with $N_H$ fixed to that of super-hot model
(green). \label{fig_ultrahot_4}}
\end{figure}

\end{document}